\documentclass[twocolumn,amsmath,amssymb,amsfonts,floatfix,showpacs]{revtex4}
\usepackage{graphicx,color, bm} 
\usepackage{dcolumn} 
\usepackage{enumerate}
\usepackage[normalem]{ulem}

\DeclareMathOperator*{\argmax}{arg\,max}

\begin{document}
\title{Bayesian inversion for the filtered flow at the Earth's Core Mantle Boundary}
\author{J. Baerenzung$^{1}$, M. Holschneider$^{1}$, V. Lesur$^{2}$}
\affiliation{$^1$ Institute for Mathematics, University of Potsdam, Potsdam, Germany \\
$^2$ Helmholtz Center Potsdam, GFZ German Research Centre for Geosciences,
Potsdam, Germany}
\begin{abstract}
The inverse problem which consists of determining the flow at the Earth's Core Mantle Boundary
according to an outer core magnetic field and secular variation model, has been investigated
through a Bayesian formalism. 
To circumvent the issue arising from the truncated nature of the available fields, we combined two modelization methods.
In the first step, we applied a filter on the magnetic field to isolate its large scales by reducing the
energy contained in its small scales, we then derived the dynamical equation, referred as filtered frozen flux equation, describing 
the spatio-temporal evolution of the filtered part of the field. 
In the second step, we proposed a statistical parametrization of the filtered
magnetic field in order to account for both its remaining unresolved scales and its large scale uncertainties.
These two modelization techniques were then included in the Bayesian formulation of the inverse problem.
To explore the complex posterior distribution of the velocity field resulting from this development,
we numerically implemented an algorithm based on Markov Chain Monte Carlo methods. After evaluating our approach
on synthetic data and comparing it to previously introduced methods, we applied
it on real data for the single epoch $2005.0$. 
\end{abstract}
\maketitle

\section{Introduction}
The core magnetic field (MF) of the Earth is sustained by the dynamo action taking place in its outer core. 
Here, the variations in chemical composition and in temperature of the liquid metal allows
convection to develop. Since the fluid is electrically conducting, it interacts non-linearly with the magnetic field. 
While the flow is advecting the MF, this latter is constraining the fluid motions through 
the Lorentz forces. The evolution of the fluid velocity field (VF) and the magnetic field are therefore entirely 
connected to one another through energy exchanges between them.

\noindent Studying such a system is difficult in many aspects. From a numerical point of
view, simulating directly the dynamic of the outer core is a challenging task. 
Because of the strong regime of turbulence that
the magnetohydrodynamic (MHD) flow exhibits, the separation between the smallest 
and the largest scales of the system is extremely broad. Yet, to properly describe the evolution
of both the VF and the MF, all these scales should be considered in simulations since they interact non-linearly together.
But this is, with the actual computation power available, impossible.

\noindent From an experimental point of view, observing directly the evolution of the outer core is also impossible.
Nevertheless, measurements of the MF from ground observatories or satellites may allow to indirectly
infer some dynamical properties of the flow and the MF in the core. 
In particular, because of the low conductivity of the mantle (\cite{Velimsky2010}), a knowledge
of the core MF at the Earth's surface is sufficient to evaluate it at the level
of the core-mantle boundary (CMB). At this very specific location, the MF is coupled to the outer core VF 
through the frozen flux equation (a simplified version of the induction equation introduced by \cite{Roberts1965}, and in
which the diffusion effects are neglected).
By inverting this equation it is therefore possible to evaluate the VF at the CMB. Unfortunately, the problem
is ill-posed for different reasons. First, the two components of the velocity field are connected to the
radial component of the secular variation (SV) through a unique equation. Then, the available secular variation
given by MF models such as GRIMM 2 of \cite{Lesur2010} is only resolved at large scales 
whereas any scale of the velocity field can contribute to this resolved SV. 
Finally, as it is the case for the SV, only the large scale core MF can be determined at the Earth surface, 
yet interactions between the unknown small scale MF with the VF can generate large scale SV (see \cite{Eymin2005}).

\noindent As shown by \cite{Backus1968}, to resolve the non-uniqueness of the velocity field in this inverse problem, 
constraints have to be imposed to the flow behavior. Different formulations have been
proposed over the past decades. This includes tangential-geostrophy, tangential-magnetostrophy, columnar flow, helical flow,  or purely toroidal flow
(see \cite{Holme2007} and \cite{Finlay2010} for an exhaustive review of the different constraints usually applied and their 
physical implications). Nevertheless, these physical constraints are not sufficient to 
provide a unique flow solution (see \cite{Chulliat2000} and references 
therein), and additional regularization assumptions have to be introduced in the problem (see \cite{Holme2007}).  

\noindent Although identified for more than twenty years (\cite{Hulot1992}), it is only recently that the effects
of the unknown small scale MF on the large scale SV have been modeled in the inversion of the Frozen Flux approximation.
In particular two methods have given promising results, namely the ensemble approach of \cite{Gillet2009}, and the iterative
method of \cite{Pais2008}. The philosophy of these two approaches is quite distinct from one another. Whereas 
in the method  of \cite{Gillet2009} an ensemble of magnetic field containing small scales is generated
and directly used to evaluate an ensemble of velocity field, in the method of \cite{Pais2008} the effects of the unknown magnetic 
field is transposed into a modeling error which is iteratively estimated. In this study, we propose 
a development of these approaches in the context of Bayesian modeling. The article is therefore constructed in the following manner.

\noindent In section \ref{governingEquations}, after describing the principle of derivation of the Frozen Flux equation, we recall the issue raised in the inversion 
of this equation by the truncated nature of the 
available MF. We then describe how to isolate the large scales of the MF and present the approximated dynamical equation,
referred as Filtered Frozen Flux (FFF) equation, 
which determines the evolution of these large scales. At the end of the section we present how to formulate the
inverse problem in a Bayesian framework when variations of the MF around the prescribed one are allowed to occur.  In section
\ref{numericalMethod} tests on synthetic data are performed. In the first one, we evaluate the improvement brought by using the FFF equation
in the inverse problem. In the second test, the Bayesian formalism developed in section \ref{governingEquations}
is considered to recover the velocity field from a set of artificially generated data, and the results are compared to the flow
obtained with three other approaches.
The methodology we developed is then used to determine the velocity field and its underlying uncertainties for the epoch $2005.0$.
Finally we present our conclusions in section \ref{conclusion}.

\section{Governing equations}\label{governingEquations}

\subsection{The Frozen-Flux approximation}
In this section we introduce briefly the hypothesis necessary to derive the
Frozen Flux approximation. For a more detailed description see \cite{Holme2007}.

\noindent Outside the core, the geodynamo's MF $\bf B$ is irrotational, it can therefore
be expressed through a potential $\phi$ according to the relation:
\begin{equation}\label{potential}
{\bf B} = -{\bf \nabla} \phi \ .
\end{equation}
As mentioned previously, the low conductivity of the mantle (\cite{Velimsky2010}) allows
to evaluate the MF and therefore its radial component $B_r = -\partial_r \phi$ at the level of the core 
mantle boundary. There its evolution is prescribed by the induction equation:
\begin{equation}
\label{CMBinduction}
\partial_t B_r = - {\bf \nabla_{_H} }({\bf u}{B}_r) + \eta \left(\Delta {\bf B} \right)\cdot {\bf e_r} \ ,
\end{equation}
with ${\bf \nabla_{H} }$ the horizontal divergence operator, ${\bf u}$ the two-dimensional
velocity field, $\eta$ the magnetic diffusivity, and ${\bf e_r}$ a radial unitary vector.
\noindent Because for the Earth, on short period of time the dissipation
effects are dominated by advection effects (see \cite{Holme2007})
and since our study is limited to single epoch inversion, the
fluid can be considered as a perfect conductor. \cite{Roberts1965} showed that
under this assumption, known as the Frozen Flux (FF) approximation, the induction
equation can be simplified as:
\begin{equation}\label{frozenflux}
\partial_t {B}_r = - {\bf \nabla_{_H} }({\bf u}{B}_r) \ .
\end{equation}

\subsection{The unresolved scale issue}\label{TUSI}

To perform a consistent inversion of the Frozen-Flux equation at the CMB, 
in addition to imposing a certain behavior to the flow and to regularizing it (see \cite{Holme2007,Finlay2010}), every
scale composing the secular variation and the magnetic field  have to be known. Unfortunately,
in the actual models describing the spatio-temporal evolution of the Earth's magnetic
field derived from satellite and observatory data, the resolution of both the SV and the MF
is limited. For the GRIMM 2 model of {\cite{Lesur2010}}, for example, the fields do not
excess the degree $13$ when expanded in spherical harmonics. 
So if $g_{l,m}$ corresponds to the spherical harmonics (SH) coefficient 
at degree $l$ and order $m$ associated with the scalar potential $\phi$,
such as:
\begin{equation}
\phi = R \sum_{l=0}^{l=+\infty} \sum_{m=-l}^{m=+l}
\left(\frac{c}{r} \right)^{l+1}g_{l,m}Y_{l,m} \ , 
\label{SHdecompostion}
\end{equation}
where $R$ is the core radius, and $Y_{l,m}$ is the Schmidt semi-normalized SH, then according to 
equation  (\ref{potential}), the available MF $B_r^<$, and its unknown part $B_r^>$ are respectively
given by:
\begin{eqnarray}
B_r^< & = & -\sum_{l=1}^{l=lc} (l+1)\sum_{m=-l}^{m=+l}
g_{l,m}^<Y_{l,m} \\
B_r^> & = & -\sum_{l=lc+1}^{l=+\infty} (l+1)\sum_{m=-l}^{m=+l}
g_{l,m}^>Y_{l,m} \  , 
\label{BrSHdecompostion}
\end{eqnarray}
with $lc$ the cut-off scale (which is equal to $13$ for the MF provided by the GRIMM 2 model). 
Note that the total radial component of the MF corresponds to the 
sum of these two quantities.

\noindent To account for the truncated nature of the available MF and SV, the Frozen Flux approximation 
has to be rewritten as:
\begin{eqnarray}
\partial_t {B}_r^< & = & - \left( {\bf \nabla_{_H} }({\bf u}{B}_r)\right)^< \label{truncatedFF}\\
& = & - \left( {\bf \nabla_{_H} }\left({\bf u}{B}_r^<\right)\right)^<
 - \left( {\bf \nabla_{_H} }({\bf u}{B}_r^>)\right)^< \label{truncatedFF1} \ ,
\end{eqnarray}
where the advection term, on the right hand side of equation (\ref{truncatedFF1}), 
is decomposed into two parts, one depending on the resolved magnetic field and the
other function of the undetermined field. \cite{Hulot1992} were the first to highlight the
issue raised by the unknown part of the MF in the inversion of the FF equation. Nevertheless,
because at this epoch the uncertainties on SV measurements were large, the contribution
of the term $\left( {\bf \nabla_{_H} }({\bf u}{B}_r^>)\right)^<$ could be neglected in the inverse
problem. The recent increase in quality of both the measurements and models describing the evolution of the core MF
(see \cite{Hulot2002}) invalidate this latter statement, and \cite{Eymin2005}
showed that the unresolved part of the MF could not be neglected anymore.

\subsection{Parametrization of the unresolved magnetic field}\label{extrapolation}

To model the effects of the unresolved part of the magnetic field on the large scale secular variation,
assumptions on this unknown field behavior have to be made. In particular one can prescribe to it a certain
energy spectrum. This operation can be performed, for instance, by extrapolating the spectrum 
associated with the resolved scales which is defined by:
\begin{equation}
E_{B^<}(l) = (l+1)\sum_{m=-l}^{m=l} \left(g_{l,m}^<\right)^2 \ . \label{SMF} 
\end{equation}
The resulting spectrum $E_{B^>}$ can then be used to statistically model the unknown MF $B_r^>$.
Following the development of \cite{Hulot1992} where the field is assumed to be isotropically distributed, 
the covariance of the coefficients $g_{l,m}^>$ is directly proportional to the extrapolated spectrum through the relation:
\begin{equation}
E[g^>_{l,m},g^>_{l^\prime,m^\prime} ] = \frac {E_{B^>}(l)}{(l+1)(2l+1)}\delta(l-l^\prime) \delta(m-m^\prime) \ , \label{variances}
\end{equation}
where $E[...]$ corresponds to the mathematical expectation.

\begin{figure}[h!]
\begin{center}
\includegraphics[width=\linewidth]{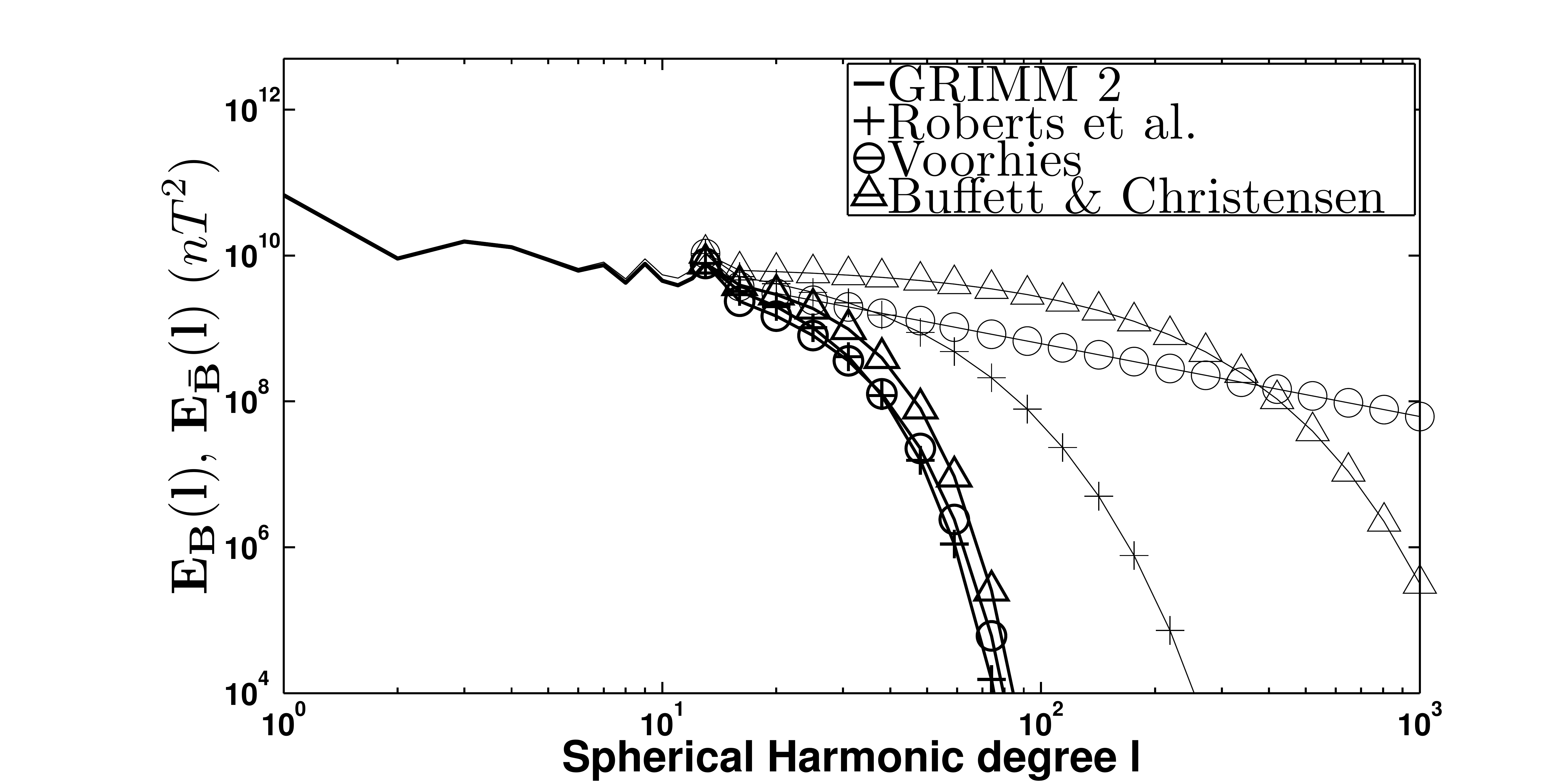}
\caption{Energy spectrum of the GRIMM 2 magnetic field at the CMB for the epoch $2005.0$ (between SH degree $1\leq l \leq 13$)
and its filtered (thick lines) and non filtered (thin lines) extrapolations. The laws used to extrapolate the large scale
spectrum were taken from  \cite{Roberts2003} (plus symbols), \cite{Voohries2004} (circles) and \cite{Buffett2007} 
(triangles).}
\label{filteredExtendedSpectra}
\end{center}
\end{figure}

\noindent The issue with such a modelization is that no universal spectrum of the geodynamo's MF at the core mantle boundary 
is available. Nevertheless, different formulations have been proposed over the past decades, but unfortunately most of them strongly 
differ from one another. An illustration of this statement is given in figure \ref{filteredExtendedSpectra} where the spectrum of the MF prescribed 
by the GRIMM 2 model for the epoch $2005.0$ (between SH degree $1\leq l \leq 13$) together with three different extrapolations (thin lines)
are plotted.
In this example, the laws derived by \cite{Buffett2007}, \cite{Roberts2003} and \cite{Voohries2004} were used to extrapolate
the resolved scale spectrum. They respectively read:
\begin{eqnarray}
E_B^B(l) &=& C_1 \chi^l \label{buffett} \\
E_B^R(l) &=& C_2 e^{-Sl} \label{roberts} \\
E_B^V(l) &=& C_3 \frac{l+1/2}{l(l+1)} \label{voohries} \,
\end{eqnarray}
with $\chi=0.99$ and $S=0.055$. The constants $C_1$, $C_2$ and $C_3$ were determined by fitting the resolved magnetic 
field spectrum between the degree $2 < l \leq 13$.
\noindent As it is confirmed in figure \ref{filteredExtendedSpectra} the three extrapolated spectra are presenting distinct behaviors.
Furthermore, the law proposed by \cite{Buffett2007} predicts a strong concentration of energy at small scales. If this 
latter formulation was to be the closest one from reality, the unresolved MF would have to be modeled at very high degree (up to $l\sim500$), which
is nowadays numerically impossible.
We propose therefore to reduce the impact of the small scale magnetic field on the large scale secular variation by
applying a filter on the magnetic field and by determining the dynamical equation governing the evolution of the resulting filtered field.

\subsection{Filtering of the fields}

As it is the case for the Earth's dynamo, the degree of turbulence in astrophysical 
or geophysical systems is most of the time extremely high. This implies that the
fields are populated by a great variety of scales. Capturing all these scales
in observations or in numerical simulations is usually impossible. 
Therefore, focusing on the large scales,
and their dynamics, is probably the best solution to study such systems. This approach
has been widely followed in the context of hydrodynamic (HD) and magnetohydrodynamic (MHD) 
turbulence (see \cite{Sagaut2006,Lesieur2008,Baerenzung2008a,Baerenzung2008b,Fabre2011}).
To extract the large scales from a given field, one has to filter it. 
In HD or MHD turbulence three different filters are generally used, the top-hat or the
Gaussian filter for studies in Cartesian space, or the sharp cut-off filter in spectral space.
While the top-hat or Gaussian filter are continuous filters, the cut-off filter truncate
the field above a certain predefined scale.
In spherical coordinates the top-hat filter is usually preferred (\cite{Sun2007}) since until
recently no equivalent of the Gaussian filter existed for such geometries (see however 
\cite{Bulow2004} ).
In our study, we decided to consider the isotropic filter developed by \cite{Bulow2004}.
The principle of derivation of this filter is detailed in appendix \ref{appA}. 
\noindent Applying this filter to a scalar field ${\xi}$ leads to:
\begin{equation}\label{convolution}
\overline{\xi} = G\star \xi \ ,
\end{equation}
where $\overline{\xi}$ is the filtered field, the $\star$ symbol denotes 
the convolution product over the sphere between convolution kernel $G$ and the scalar field $\xi$.
According to the convolution theorem (see \cite{Driscoll1994}), in spectral space equation (\ref{convolution}) becomes:
\begin{equation}\label{spectralConvolution}
\overline{\xi}_{l,m} =  \sqrt{\frac{4\pi}{2l+1}} \xi_{l,m} G_l^0 \ ,
\end{equation}
where $\overline{\xi}_{l,m}$ and $\xi_{l,m}$ are respectively the filtered and the total
SH coefficients associated with $\xi$, and $G_l^0$ is convolution
kernel expressed in spectral space at degree $l$. 
Since for the filter developed by \cite{Bulow2004}, $G$ is zonal, 
only its coefficients at order $m=0$ are different from zero.
As shown in appendix \ref{appA} equation (\ref{spectralFilter}), equation 
(\ref{spectralConvolution}) can be expressed as:
\begin{equation}\label{spectralConvolution1}
\overline{\xi}_{l,m} = \xi_{l,m}  \exp{\left( -\frac{l(l+1)\overline{\Delta}^2}{24 R^2}\right)} \ .
\end{equation}
where $\overline{\Delta}$ corresponds to the filter width.
With this formulation one can directly connect the spectrum of the scalar field $E_\xi$
to the spectrum of its filtered part $E_{\overline{\xi}}$ as:
\begin{equation}\label{filteredSpectrum}
E_{\overline{\xi}} = E_\xi  \exp{\left( -\frac{l(l+1)\overline{\Delta}^2}{12 R^2}\right)} \ .
\end{equation}

\noindent We then applied this filter to the three different extrapolated spectra presented in the previous
section (see equations (\ref{buffett})-(\ref{voohries})).
In figure \ref{filteredExtendedSpectra} the non filtered
(thin lines), and the filtered (thick lines) spectra are plotted. The  
width of the filter has been set to $\overline{\Delta}=500$km in order to preserve the large scale
fields and to suppress the small scale ones. One can observe that most of the energy 
that was contained in the small scales of the initial fields has now vanished in the
filtered fields. Furthermore, the variations between the different extrapolated
filtered spectra are now much less pronounced than in the non filtered case.

\subsection{The Filtered Frozen-Flux (FFF) approximation }\label{FFequations}

Applying the spatial filter presented in the preceding section to the velocity and 
the magnetic field allows to decompose them into an averaged part 
$\overline{{\bf u}}$ and $\overline{B}_r$, and a fluctuating part ${\bf u}^\prime$ 
and $B_r^\prime$ such as:
\begin{eqnarray}
{\bf u} & = & \overline{{\bf u}} + {\bf u}^\prime \\
B_r & = & \overline{B}_r + B_r^\prime \ .
\end{eqnarray}
Since the spatial filter considered here is homogeneous and time invariant, it commutes
with all the differential operators encountered in the induction equation. 
Therefore applying this filter to the Frozen Flux approximation
(equation (\ref{frozenflux})) leads to:
\begin{equation}\label{filteredfrozenflux}
\partial_t \overline{B}_r = - {\bf \nabla_{_H} }(\overline{{\bf u}{B}}_r) \ .
\end{equation}

\noindent Following the decomposition introduced by \cite{Leonard1974}, the part of the subgrid
stress tensor which allows interactions between the tangential components of the velocity
field and the radial component of the MF reads:
\begin{equation}\label{subgridstress}
{\bf \tau} = \overline{{\bf u}{B}}_r - \overline{{\bf u}} \overline{B}_r \ .
\end{equation}
Now equation (\ref{filteredfrozenflux}) can be rewritten as:
\begin{equation}\label{FFF}
\partial_t \overline{B}_r = - {\bf \nabla_{_H} }(\overline{{\bf u}} \overline{B}_r )
- {\bf \nabla_{_H} } {\bf \tau} \ .
\end{equation}
The subgrid stress tensor $\tau$, through the averaged product $\overline{{\bf u}{B}}_r
=\overline{(\overline{\bf u}+u^\prime)(\overline{B}_r + B_r^\prime)}$,
incorporates interactions between averaged and fluctuating part of the fields, 
thus equation (\ref{FFF}) is not closed in the sense that it does not only contains large
scale quantities.
To close this equation, expressions of the fluctuating quantities depending on the
averaged ones have to be derived. 

\noindent As mentioned in appendix \ref{appA},
the filtered velocity and magnetic field are both solutions of the diffusion equations:
\begin{eqnarray}
\partial_s {\bf u}({\bf x},s) & = & D_u \Delta_{_H} {\bf u}({\bf x},s) \label{diffusionU}\\
\partial_s B_r({\bf x},s) & = & D_{_B} \Delta_{_H} B_r({\bf x},s) \label{diffusionB} \\
\partial_s \left({\bf u}({\bf x},s)B_r({\bf x},s)\right) & = & D_{_B} \Delta_{_H} 
\left({\bf u}({\bf x},s)B_r({\bf x},s)\right)\label{diffusionUB}
\end{eqnarray}
where the filtered and total fields are respectively $\overline{\bf u}={\bf u}({\bf x},s)$ 
and $\overline{B}_r = B_r({\bf x},s)$, and ${\bf u}={\bf u}({\bf x},0)$ and $B_r=B_r({\bf x},0)$, and
the product between $s$ and the two diffusion coefficients $D_B$ and $D_u$ determines the width of the filter.
Note that since $D_B$ is the diffusion coefficient applied to each term of the 
Frozen Flux equation it has to be applied to the MF but not necessarily to the VF. 
To link fluctuating to filtered fields, Taylor expansions at the first order in $s$ of 
${\bf u}$, $B_r$, and $\overline{{\bf u}{B}}_r$, using relations (\ref{diffusionU}) to
(\ref{diffusionUB}) are performed, leading to:
\begin{eqnarray}
{\bf u} & \sim & \overline{\bf u} - sD_u \Delta_{_H} \overline{\bf u} \label{approxU}\\
B_r & \sim & \overline{B}_r - sD_{_B}\Delta_{_H} \overline{B}_r \label{approxB} \\ 
\overline{{\bf u}B}_r & \sim &  {\bf u}B_r + sD_{_B}\Delta_{_H}  ({\bf u}B_r) \label{approxUB} .
\end{eqnarray}

\noindent By analogy to the usual Gaussian filters used in Large Eddy Simulations, one can define the
following characteristic lengths:
\begin{eqnarray}
{\overline{\Delta}}^2_u &=& 24 sD_u \label{lengthU} \\
{\overline{\Delta}}^2_{_B} &=& 24 sD_{_B} \label{lengthB}
\end{eqnarray}
Letting ${\overline{\Delta}}^2_u = {\overline{\Delta}}^2_{_B} = {\overline{\Delta}}^2$, 
injecting equations (\ref{approxU}) and (\ref{approxB}) into equation (\ref{approxUB}), and
keeping only the first order terms, one gets:

\begin{equation}\label{tau1}
\overline{{\bf u}B}_r \sim \overline{{\bf u}} \overline{B}_r 
- \frac{{{\overline{\Delta}}^2}}{24}\left( \overline{{\bf u}} 
\Delta_{_H} \overline{B}_r +\overline{B}_r \Delta_{_H} \overline{{\bf u}}
- \Delta_{_H}( \overline{{\bf u}}\overline{B}_r)\right) \ ,
\end{equation}
which can be reduced to:
\begin{equation}\label{tau}
\overline{{\bf u}B}_r \sim \overline{{\bf u}} \overline{B}_r 
+ \frac{{{\overline{\Delta}}^2}}{12} \left( \left( {\bf \nabla_{_H}} \overline{B}_r
\right) {\bf \nabla_{_H}} \right) \overline{\bf u} \ .
\end{equation}
So the subgrid stress tensor $\tau$ in its approximated closed form reads:
\begin{equation}\label{tau}
{\bf \tau} = \frac{{{\overline{\Delta}}^2}}{12} \left( \left( {\bf \nabla_{_H}} \overline{B}_r
\right) {\bf \nabla_{_H}} \right) \overline{\bf u} \ .
\end{equation}
The Filtered Frozen Flux equation can now be written for the filtered secular
variation $\partial_t \overline{B}_r^<$ truncated at degree $l_c=13$ as:
\begin{equation}\label{TFFF}
\partial_t \overline{B}_r^< = - \left({\bf \nabla_{_H} }(\overline{{\bf u}} \overline{B}_r )
\right)^< - \left({\bf \nabla_{_H} } {\bf \tau}\right)^< \ ,
\end{equation}
with $\overline{B}_r = \overline{B}_r^< + \overline{B}_r^>$.

\noindent In our study, the extrapolated filtered magnetic field $\overline{B}_r^>$
is extended up to the spherical harmonic degree $l=30$. Above this scale, the filtered magnetic 
field (for $\overline{\Delta} = 500$ km) is extremely weak (see figure \ref{filteredExtendedSpectra}) 
and its influence on the large scale secular variation is neglected.

\subsection{Bayesian formulation of the inverse problem}

The problem of determining the velocity field at the CMB knowing the exact
MF and the SV together with its uncertainties is an ill-posed inverse problem: in a discrete approximation, the 
number of unknown is twice as large as the number of equations. One of the most
common methods to tackle this problem consists in minimizing an energy functional composed
of two main terms; a quantity measuring the discrepancies between the model and the data,
balanced, through a regularization parameter, with a quantity expressing a prior
knowledge on the expected solution. This  method has been widely used to evaluate the flow at the CMB and 
for a review of the different parametrization employed see \cite{Holme2007}. 

\noindent Another option consists in formulating the problem in a Bayesian framework.
The solution becomes then the full posterior distribution of the velocity field
given the secular variation. This method allows the estimation of a model for the flow together with 
the quantification of its uncertainties. 

\noindent When variations around the prescribed MF are allowed to occur, the inverse problem becomes more complicated.
For models of the Earth's core MF derived from satellite or observatory data, the nature of these variations is diverse. 
At large scales they can be due to a leakage of the external and lithospheric field into the core field, whereas at small scales, the entire MF
is undetermined because of the dominance of the lithospheric field at the Earth's surface. Recent models
such as GRIMM 2 (\cite{Lesur2010}) are able to separate at large scales (between SH degree $1\leq l \leq 13$)
the external from the core field, but not the lithospheric field from the core field. As a consequence, the large scale lithospheric 
field becomes a source of uncertainty on the geodynamo's MF.

\noindent To parametrize the unresolved part of the MF in the inverse problem different approaches have been recently developed. This includes
the ensemble method of \cite{Gillet2009} and the iterative algorithm of \cite{Pais2008}. In this study we propose
to extend these methods to the context of Bayesian modeling, following the development of \cite{Jackson1995}. 
Furthermore, in addition to parametrizing the unresolved MF, we also consider the uncertainties on the large scale MF due to the lithospheric
field.

\subsubsection{CMB velocity distribution}\label{baysian}

From now on, in order to simplify the notations, the filtered  MF and VF as well
as the filtered and truncated SV will be written as:
\begin{eqnarray}
\overline{B}_r & = & b \\
\overline{\bf u} & = & u \\
\dot{\overline{B}}_r^{<} & = & \gamma
\end{eqnarray}

\noindent In this section, the distribution we want to characterize is the posterior distribution
of the velocity field given the secular variation $p\left(u |\gamma \right)$.
But since we want to account for the unknown small scale magnetic field together with
the uncertainties on the large scale field, this distribution cannot be expressed directly.
Nevertheless, it can be obtained by marginalizing the joint posterior distribution of the velocity field and
the magnetic field as following:
\begin{equation}
p\left(u |\gamma \right) = \int{ p(u,b|\gamma) \mathrm{d}b}\ . \label{posterior1}
\end{equation}

\noindent According to \cite{Bayes1763}  the distribution on the right hand side of relation (\ref{posterior1}) 
can be decomposed into:
\begin{equation}
p(u , b | \gamma) = 
\frac{p ( \gamma | u,b )p\left(u,b \right)}
{p (\gamma) }\label{posteriorTot} \ ,
\end{equation}
with $p\left(u,b \right)$ the joint prior distribution of the VF and the MF,
$p (\gamma)$ the distribution of the SV, which is constant with respect to both $u$ and $b$, and finally
$p ( \gamma | u ,b)$ the likelihood distribution. Because $u$ and $b$ are a priori assumed to be independent 
random variables their joint distribution can be split into two distributions such as:
\begin{equation}\label{jointUB}
p\left(u,b \right) = p\left(u \right)p\left(b \right) \ .
\end{equation}
The prior distribution of the VF $p(u)$ is assumed to be Gaussian with the following general form:
\begin{equation}\label{priorU}
p(u) = \frac{exp\left[-\frac{1}{2}u^T \Sigma_u^{-1} u \right]}{(2\pi)^{\frac{d}{2}} 
|\Sigma_u|^\frac{1}{2}}\,
\end{equation}
with $d$ the dimension of the VF vector, which in our case is twice as large as the
dimension of the MF and SV, and $\Sigma_u$ the velocity covariance 
matrix chosen to enforce the spatial smoothness of the flow. For this purpose we impose that:
\begin{eqnarray}\label{norm}
u^T \Sigma_u^{-1} u & = &
\int_{\Omega} \left( |{\bf \nabla_{_H}}\left({\bf \nabla_{_H}}\cdot
u\right)|^2 + |{\bf \nabla_{_H}}\left( 
{\bf r}\times{\bf \nabla_{_H}}\cdot u \right) |^2 \right)\mathrm{d}\omega \nonumber \\
&  + & \alpha^2 \int_{\Omega}|u|^2 \mathrm{d}\omega \ .
\end{eqnarray}
The right hand-side of equation (\ref{norm}) is composed of two different
norms on $u$, namely the Bloxham's "strong norm" (\cite{Bloxham1988,Jackson1993}) for the 
first one and the standard $L^2$-norm for the second one. The domain of integration
$\Omega$ is the surface of the sphere describing the CMB, 
and the balance factor $\alpha^2$ is chosen to be small enough not to modify the correlation length induced
by the Bloxham norm. Furthermore, the covariance is rescaled
such as the averaged standard deviation of the velocity field intensity is $20$km/yr.
This implies that at any location of the core-mantle boundary, the probability for the 
flow intensity to excess the value of $50$km/yr, an upper limit calculated by \cite{Finlay2011},
is of the order of $0.01$.

\noindent As for the velocity field, the magnetic field $b$ is also assumed to be normally distributed,
but with a mean $b_0$ corresponding to the resolved MF, and a covariance $\Sigma_{b}$. Its
prior distribution can therefore be expressed as:
\begin{equation}
 p ( b ) =  
\frac{\exp\left[-\frac{1}{2}\left( b -  b_0 \right)^T 
\Sigma_{b}^{-1}
\left(b -  b_0  \right) \right]}{(2\pi)^{\frac{d}{4}}|\Sigma_{b}|^{\frac{1}{2}}}
\label{Bprior} \ .
\end{equation}
In this equation the two parts of the MF  $b = b^< + b^>$ 
have different behaviors. The truncated field $b^<$ is taken from the GRIMM 2 model
with $E[b^<] = b_0$ whereas the unknown MF $b^>$ is characterized by the averaged value $E[b^>] = 0$. 
Since in the GRIMM 2 model, the core field and the litospheric field 
are overlapping at large scales, this latter field can be viewed as a source
of uncertainties on $b^<$. We propose therefore to use the theoretical spectrum
of the litospheric field given by \cite{Thebault2013} to build the covariance
matrix $\Sigma_{b^<}$ for the resolved scales magnetic field.
This spectrum reads:
\begin{equation}
E_B^L(l) =  \left(l+1\right) \left( \mu_0 |M| F_l^a\left(\epsilon\right) \right)^2 l^{-\delta}C_l
\label{litosphericSpectrum} 
\end{equation}
with $|M| = 0.4225 A.m^{-1}$ the averaged crust magnetization, 
$\epsilon = 27$km the equivalent magnetized layer thickness, and
the constants $\mu_0=4\pi 10^{-7}$, $\delta=1.28$, and $a=6371.2$km. The two functions
$F_l^a\left(\epsilon\right)$ and $C_l$ are given by:
\begin{eqnarray}
C_l &=& \frac{l(l+1)(160l^5+264l^4-192l^3-130l^2+96l-9)}
{6(2l+3)^2(2l+1)^2(2l-1)^2} \nonumber \\
 F_l^a(\epsilon)& = &\frac{1-(1-\epsilon/a)^{(l-1)}}{l-1} \nonumber \ .
\end{eqnarray}
Letting the truncated MF $b^<$ being linked to its spectral counterpart $g_{l,m}^<$ through
the relation $b^< = FP g_{l,m}^<$, where the operator $P$ projects the coefficients 
in physical space, and the operator $F$ filters the field, and assuming that the litospheric
field is isotropically distributed, the covariance of the truncated MF is:
\begin{eqnarray}
\Sigma_{b^<}\! &\!\! = \!\!& \!\! E \left[\left( b^< - b_0\right)
\left( b^< - b_0\right)^T\right] \\
 &\!\! = \!\!& \!\! E \left[PF\left( g_{l,m}^< - g_{l,m}^0\right)
\left( g_{l,m}^< - g_{l,m}^0\right)^TF^TP^T\right] \\
 & \!\!= \!\!&\!\! PF \frac{E_B^L(l)}{(l+1)(2l+1)}F^TP^T \label{LSMFcovariance}\ ,
\end{eqnarray}
where $g_{l,m}^0$ are the spherical harmonics coefficients associated with $b_0$.

\noindent The extrapolated SH coefficients of the MF $g_{l,m}^>$ are assumed to individually have 
a $0$ mean and a variance depending on the extrapolated spectrum $E_{B^>}^B(l)$ from \cite{Buffett2007} 
as shown in equation (\ref{variances}). Therefore, in physical space, the MF also has
a $0$ mean at any spatial location and a covariance given by:
\begin{equation}
\Sigma_{b^>} = PF \frac{E_{B^>}^B(l)}{(l+1)(2l+1)}F^TP^T \label{SSMFcovariance}\ .
\end{equation}
By combining  $\Sigma_{b^<}$ with $\Sigma_{b^>}$, one gets the total covariance $\Sigma_{b}$ for the MF.

\noindent The last distribution to characterize is the likelihood distribution which measures the discrepancies
between the model (the Filtered Frozen Flux equation) and the data (the secular variation). It reads:
\begin{equation}
p ( \gamma|b,  u )  =  
\frac{\exp\left[-\frac{1}{2} 
\left(\gamma + A_{u }b \right)^T 
\Sigma_{\gamma}^{-1}
\left(\gamma + A_{u  }b \right)
\right] }{(2\pi)^{\frac{d}{4}}|\Sigma_{\gamma}|^{\frac{1}{2}}} \ ,\label{likelyhood}
\end{equation}
where the operator $A_{u }$, when applied to $b$,
allows to calculate the non linear term of the filtered Frozen Flux equation 
$\left({\bf \nabla_{_H} }\left(ub+\tau\right)\right)^<$, 
and the covariance matrix $\Sigma_{\gamma}$ is the SV posterior 
covariance of the GRIMM 2 model. 

\noindent All the distributions entering the velocity field posterior distribution being detailed,
this latter can be evaluated. The integral given in equation (\ref{posterior1})
has already been calculated by \cite{Jackson1995}, so we only present the result
which reads:
\begin{eqnarray}
p ( \gamma| u ) & = & \int
 p ( \gamma|b,  u ) p (b ) \mathrm{d}b \\
& \sim & \frac{(2\pi)^{\frac{d}{4}}}{|N|^{\frac{1}{2}}}\exp\left[-\frac{1}{2} 
\left(c -r^TN^{-1} r \right)\right] \label{likelihood1}
\end{eqnarray}
with:
\begin{eqnarray}
N & = &  A_{u }^T \Sigma_{\gamma}^{-1} A_{u }
+ \Sigma_{b}^{-1} \\
r & = &  \Sigma_{b}^{-1} b_0 
- A_{u }^T \Sigma_{\gamma}^{-1}\gamma \\
c & = & \gamma^T \Sigma_\gamma^{-1}\gamma
+ b_0^T \Sigma_b^{-1} b_0
\end{eqnarray}
By multiplying the density (\ref{likelihood1}) with the prior distribution (\ref{priorU})
one gets the posterior distribution: 
\begin{eqnarray}
 p(u|\gamma)  &=& 
 \frac{p ( \gamma| u ) p(u)}{p(\gamma)} \\
 &  \sim& \frac{1}{|N|^{\frac{1}{2}}}\exp\left[-\frac{1}{2} 
\left(c -r^TN^{-1} r + u^T \Sigma_u^{-1} u \right)\right]  \label{posteriorTotJ}\nonumber
\end{eqnarray}
Since this expression does not allow to easily apprehend the effects arising from the modelization of the MF variations on the 
posterior distribution of the velocity field, we decided to rewrite it into a more intuitive 
form. It reads:
\begin{eqnarray}
 p(u|\gamma)  &=& \frac{(2\pi)^{-\frac{3d}{4}}}
{  |\Sigma_{\tilde{\gamma}}|^{\frac{1}{2}}  }
\exp\left[-\frac{1}{2} 
\left(\gamma + A_{u }b_0 \right)^T 
\Sigma_{\tilde{\gamma}}^{-1}
\left(\gamma + A_{u  }b_0 \right)\right]  \nonumber \\
 & \times & \frac{1}
{ |\Sigma_u|^{\frac{1}{2}}}\exp\left[-\frac{1}{2} 
 u^T \Sigma_u^{-1} u \right]  \times  \frac{1}{p(\gamma) } \label{posteriorTot1}\\
\Sigma_{\tilde{\gamma}} & = & \Sigma_{\gamma} + A_{u} \Sigma_b A_{u}^T \ .
\end{eqnarray}
This formulation is very similar to the posterior distribution of the velocity field in the case
where the magnetic field is exactly know. Indeed this latter distribution can be obtained by simply replacing the covariance matrix $\Sigma_{\tilde{\gamma}}$ by 
$\Sigma_{\gamma}$. One can therefore observe that accounting for the small scale magnetic field
and the lithospheric field when formulating the inverse problem in a Bayesian framework leads to an increase of the secular variation uncertainties
through the quantity $A_{u} \Sigma_b A_{u}^T$.
Because of the dependency of this latter term on the velocity field $u$, the maximum of the posterior
distribution cannot be analytically calculated as already mentioned by \cite{Jackson1995}. Nevertheless, it is numerically possible to extract
the main statistical characteristics of this posterior distribution using a
Markov Chain Monte Carlo method (see \cite{Mosegaard1999,Rygaard2000}). The algorithm we chose to explore the posterior
distribution and the results we obtained are presented in the next section.

\section{Numerical method and results}\label{numericalMethod}

\begin{figure}[h!]
\begin{center}
\includegraphics[width=\linewidth]{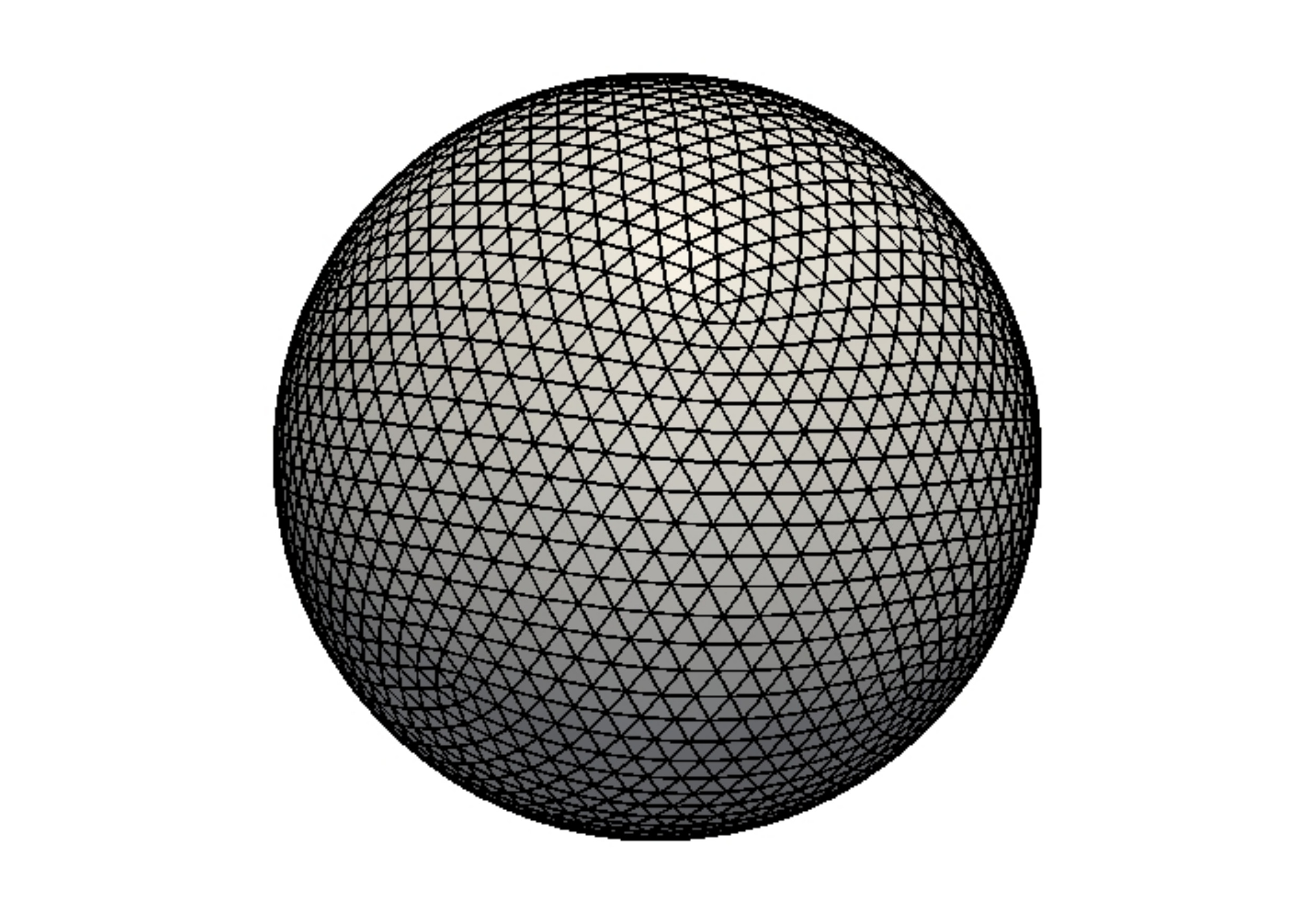}
\caption{Discrete CMB after 4 steps in the refinement of the initial icosahedron.}
\label{mesh}
\end{center}
\end{figure}

The SV, MF and VF are expressed in physical space, therefore
the surface of the CMB has been discretized by recursively dividing an initial icosahedron
(see figure \ref{mesh}). The grid construction and its properties, as well as 
the approximation of the differential operators are explicited in appendix \ref{appB}.

\noindent Given that the grid is composed of $N$ nodes, the vectors $b$, 
${\gamma}$, and ${u}$ are given by:
\begin{eqnarray}
b^T & = & \left(b_0 ... b_i...b_{N-1} \right) \ ,\nonumber \\
{\gamma}^T & = & \left(\gamma_0...\gamma_i...\gamma_{N-1}\right) \ , \nonumber \\
u^T & = & \left(u_0 ... ,u_i,...,u_{N-1}\right) \ . \nonumber 
\end{eqnarray}

\subsection{Evaluation of the Filtered Frozen flux model}\label{FFFevaluation}

To evaluate the FFF model, an inversion of this equation was performed using GRIMM 2 as well as
artificially generated data. The large scale magnetic field was taken from the GRIMM 2 
model at the epoch $2001.0$, whereas the small scales were randomly generated for SH degree 
lying between $14$ and $160$ according to the exponential law (\ref{buffett}). 
The MF was then filtered (with $\overline{\Delta}=80$km) such as its smallest scales, 
which cannot be properly represented on the grid, exhibited a low energy level.
This MF is referred as $b_0$.
To create a SV associated with this MF, we drew randomly a velocity field and used
it to advect the MF with the FF equation (\ref{frozenflux}).
The velocity field was decomposed into a poloidal and a toroidal part such as:
\begin{equation}
{\bf{u}} = {\bf{\nabla}} \Phi + {\bf{e_r}} \times {\bf\nabla} \psi \ . \label{poloToro}
\end{equation}
In spectral space, the spherical harmonics coefficients for the poloidal and toroidal
field respectively read $\Phi_{l,m}$ and $\psi_{l,m}$. In order to promote interactions between 
small and large scales, $\Phi_{l,m}$ and $\psi_{l,m}$ were extended up to spherical harmonic degree 
$80$ with the following statistical properties:
\begin{equation}
E[\Phi_{l,m}] =  E[\psi_{l,m}] = 0 \quad \forall l,m 
\end{equation}
\begin{eqnarray}
E[\Phi_{l,m} \Phi_{l^\prime,m^\prime}] &=& C l^{-14/3} |m|^{-11/3}\delta_{l l^{\prime}} 
\delta_{m m^{\prime}} \\
E[\psi_{l,m} \psi_{l^\prime,m^\prime}] &=& C  l^{-14/3} |m|^{-11/3}\delta_{l l^{\prime}} 
\delta_{m m^{\prime}} \\
E[\Phi_{l,m} \psi_{l^\prime,m^\prime}] &=& 0  \quad \forall l,l^{\prime},m ,m^{\prime}\ ,
\end{eqnarray}
where $C$ is a normalization constant. The choice of a $l^{-14/3}$ power law imposed onto
the poloidal and toroidal field correlation allowed to generate a flow with 
similar statistical properties than a two-dimensional turbulent flow (see \cite{Sukoriansky2002}).

\noindent To directly simulate the advection of the MF, a fourth order Runge-Kutta scheme had been 
implemented. The integration time was taken to be $0.05$ year, and the computation was performed on 
a grid refined $7$ times (with $163842$ nodes).

\begin{figure}[h!]
\begin{center}
\includegraphics[width=\linewidth]{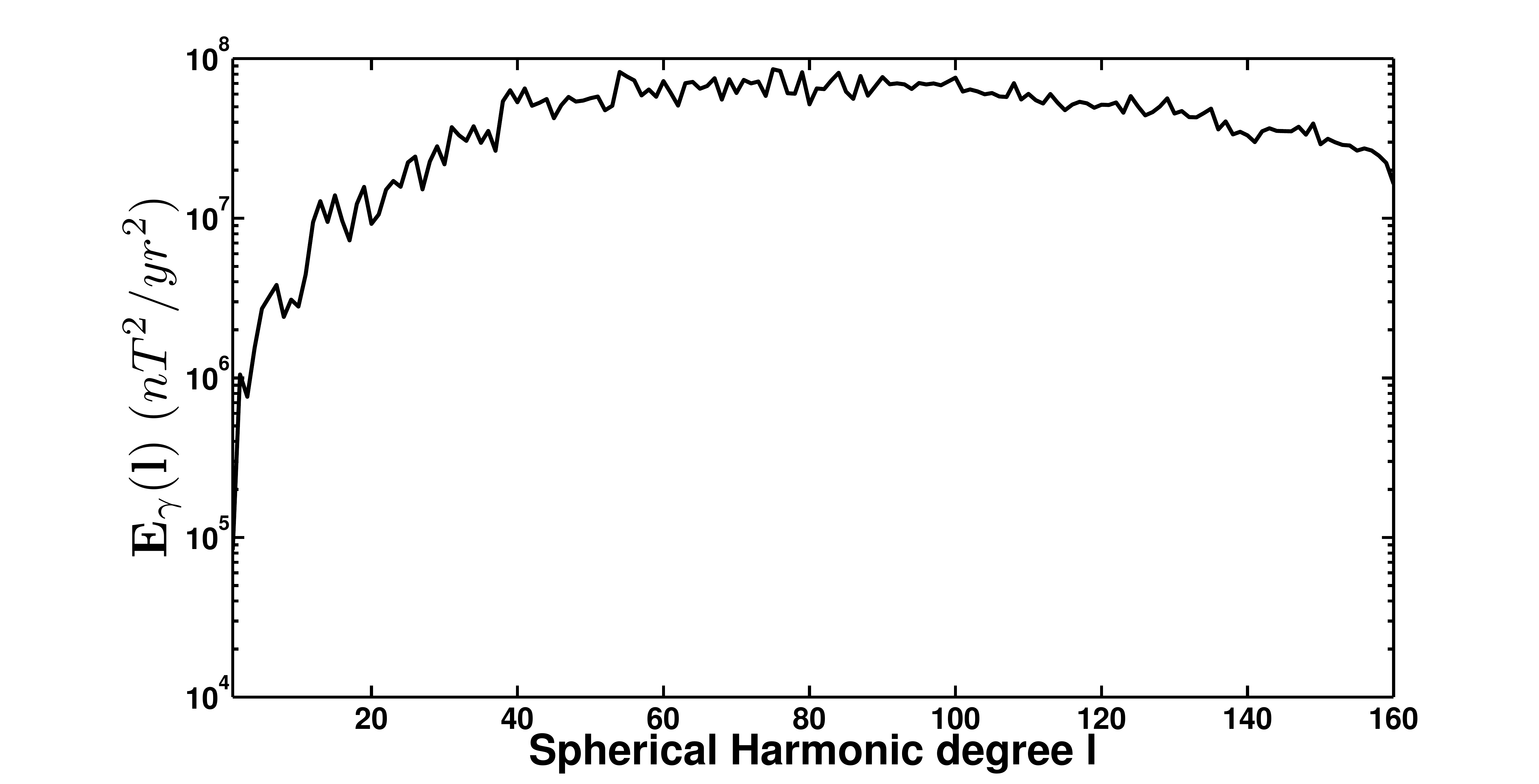}
\caption{Spectrum of the secular variation  generated by advecting the
GRIMM 2 extrapolated MF at the epoch $2001.0$ with an artificial velocity field.}\label{dtBadvected}
\end{center}
\end{figure}

\noindent In figure \ref{dtBadvected}, the spectrum of the resulting secular variation is plotted.
As it can be observed, the energy of the SV is maximum at scales lying between $l=60$ and
$l=100$ indicating that the MF is strongly affected by the velocity field at these scales.

\noindent To mimic the situation encountered when using models fitting satellite data 
where both the MF and SV cannot 
be entirely taken into account, the artificial fields were truncated at degree 40.
The resulting MF and SV are respectively referred as $\gamma$ and $b_0^<$.
These fields were then used as an input for the inversion of the FFF equation. 
Two different filter widths were tested, 
$\overline{\Delta}=500$km, and $\overline{\Delta}=0$km. In the latter case
the equation reduces then to the usual FF approximation. Since the statistical properties
of the velocity field were exactly known, they were directly injected in the prior information
$p({u})$. No measurement errors on the secular variation and the magnetic field had been 
generated, therefore the likelihood and the MF prior distributions respectively read:
\begin{eqnarray}
p\left(\gamma|u,b\right) &=& 
\delta\left(\gamma + {\bf \nabla_{_H} }(bu + \tau )\right)  \ , \\
p(b) &=& \delta(b-b_0^<) \label{artificialPriorB} \ . 
\end{eqnarray}
Multiplying together these two distributions and marginalizing the result with respect to $b$ leads to:
\begin{equation}
p\left(\gamma|u\right) =
\delta\left(\gamma + {\bf \nabla_{_H} }\left(b_0^< u + \tau\left(b_0^<\right) \right)\right) \label{artificialLikelihood} \ ,
\end{equation}
with $\tau(b_0^<)$ the subgrid stress tensor evaluated with $b_0^<$.
The posterior distribution of the velocity field is then proportional to:
\begin{equation}
p\left(u\right|\gamma) \sim p\left(\gamma|u\right)p(u) \ .
\end{equation}
The Bayesian formulation of the problem being described,
the discrete velocity field that maximizes the posterior distribution could be determined
for the two cases. To do this, a particular solution of the equation $\gamma + 
{\bf \nabla_{_H} }\left(b_0^< u + \tau\left(b_0^<\right)\right)=0$ was calculated. In addition, the null 
space of the operator $A_{b_0^<}$ defined such as 
$A_{b_0^<}u=\nabla_{_H} \left(b_0^< u + \tau\left(b_0^<\right)\right)$ was parametrized.
The final solution corresponded then to the sum of the particular solution with the null-space one
which minimized the prior information on the VF.
The grid used to realize the computation was composed of $10242$ nodes
(approximately $16$ times less than the one taken to advect the MF).
For the results to be comparable, the three different velocity fields were truncated at 
SH degree $l=40$. Furthermore the artificially generated field as well as the field
obtained by inverting the FF approximation ($\overline{\Delta}=0$),
were filtered with $\overline{\Delta}=500$km.

\begin{figure}[h!]
\begin{center}
 \begin{minipage}[c]{.49\linewidth}
      \includegraphics[width=\linewidth]{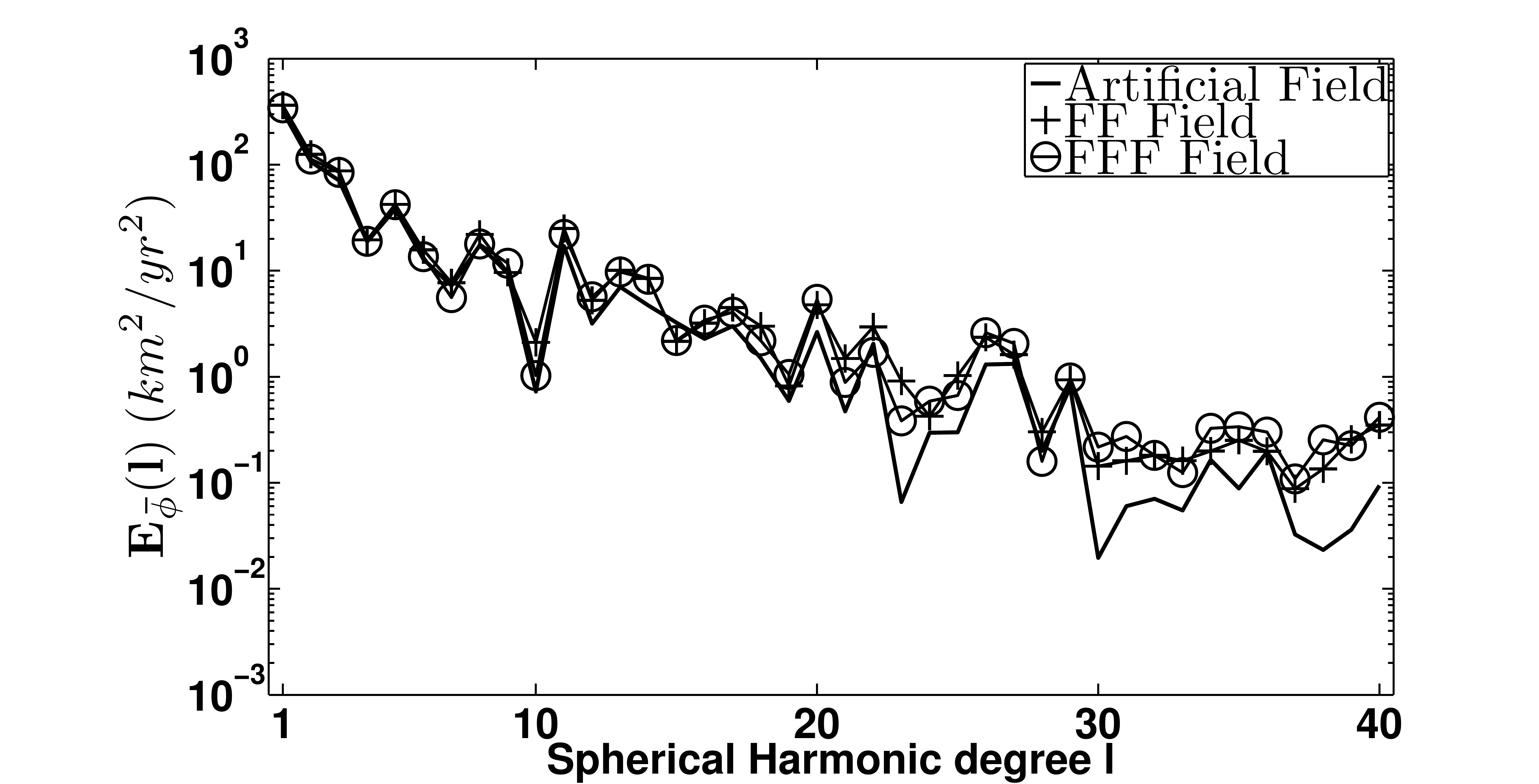}
 \end{minipage} \hfill
 \begin{minipage}[c]{.49\linewidth}
      \includegraphics[width=\linewidth]{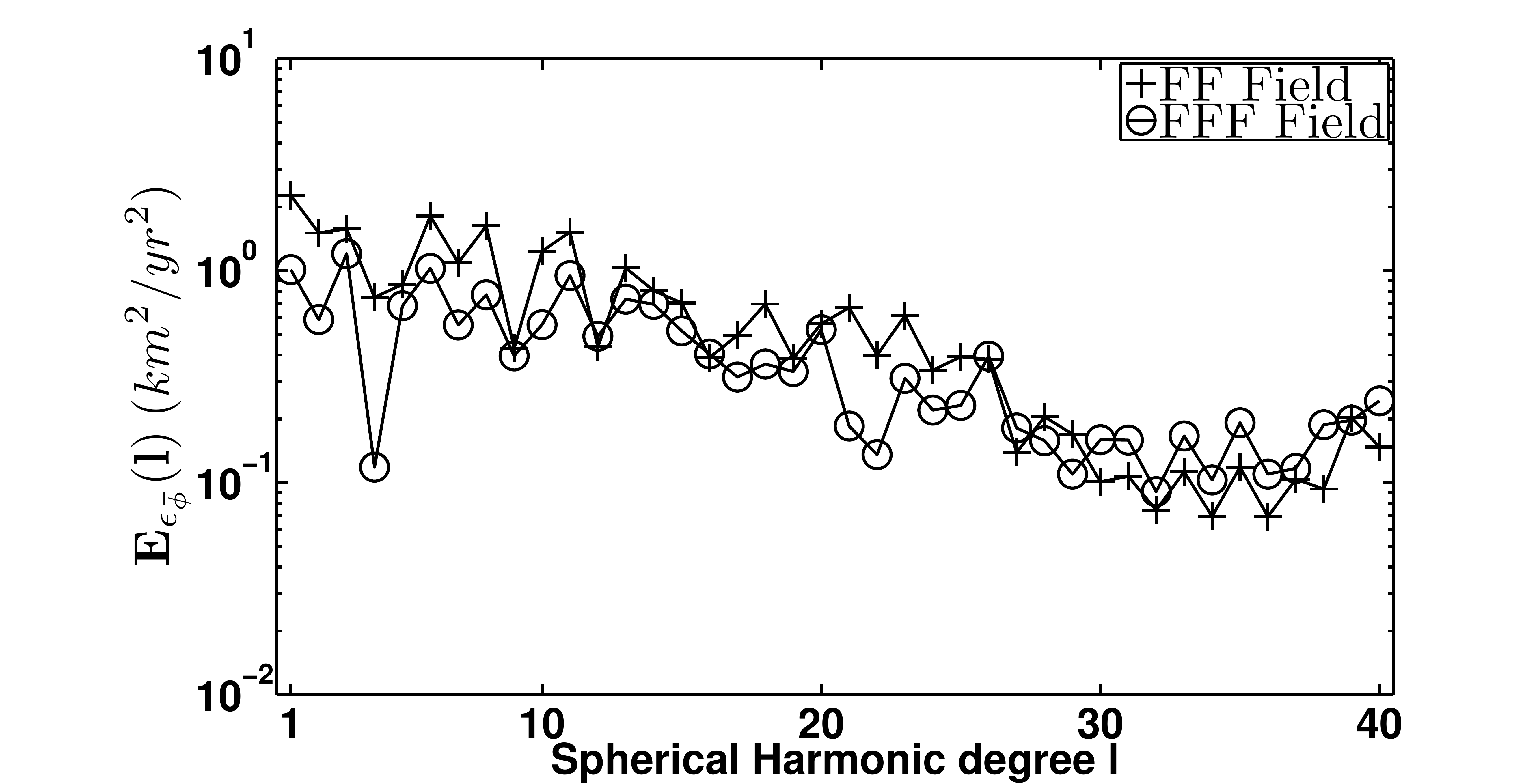}
 \end{minipage}
 \begin{minipage}[c]{.49\linewidth}
      \includegraphics[width=\linewidth]{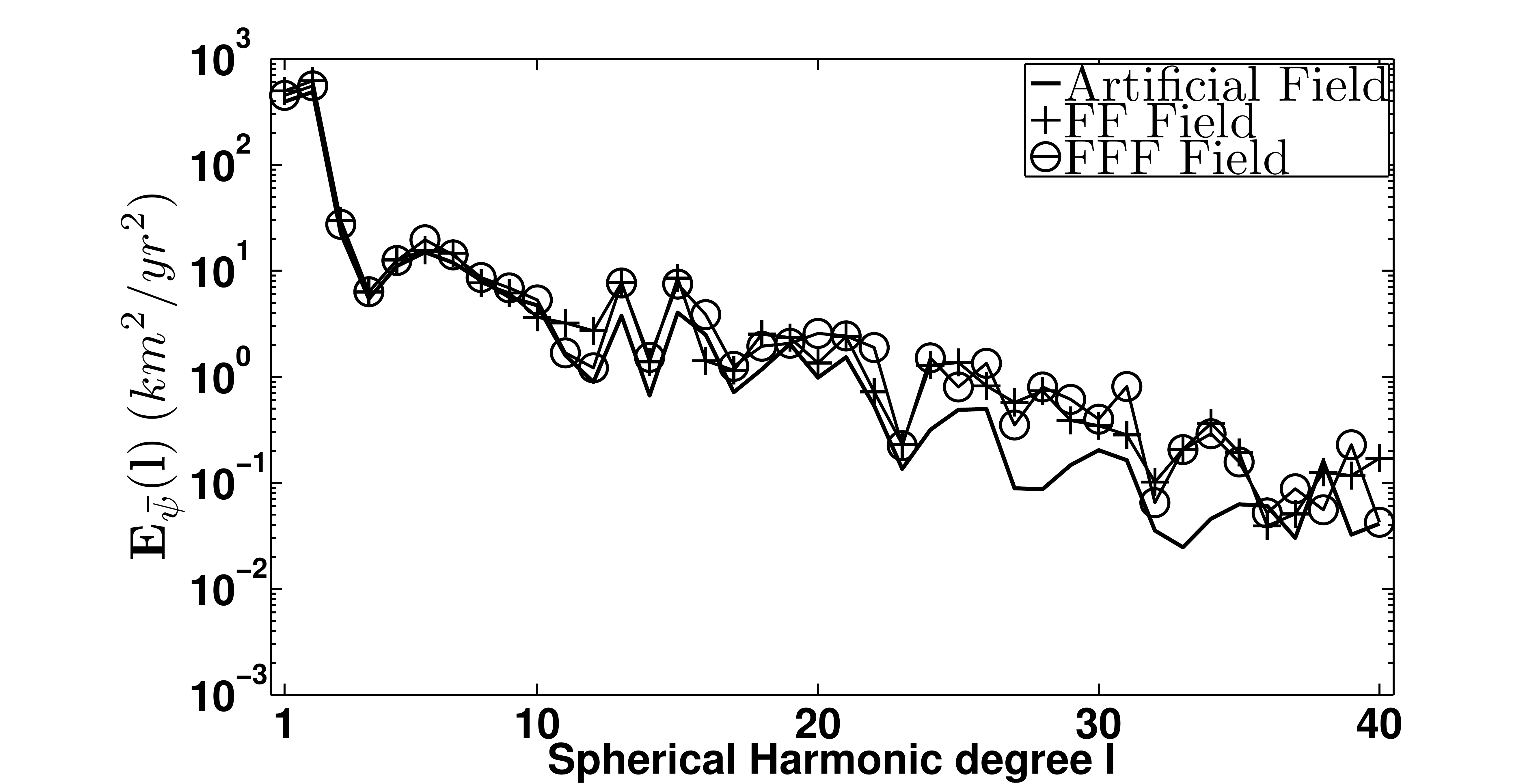}
 \end{minipage} \hfill
 \begin{minipage}[c]{.49\linewidth}
      \includegraphics[width=\linewidth]{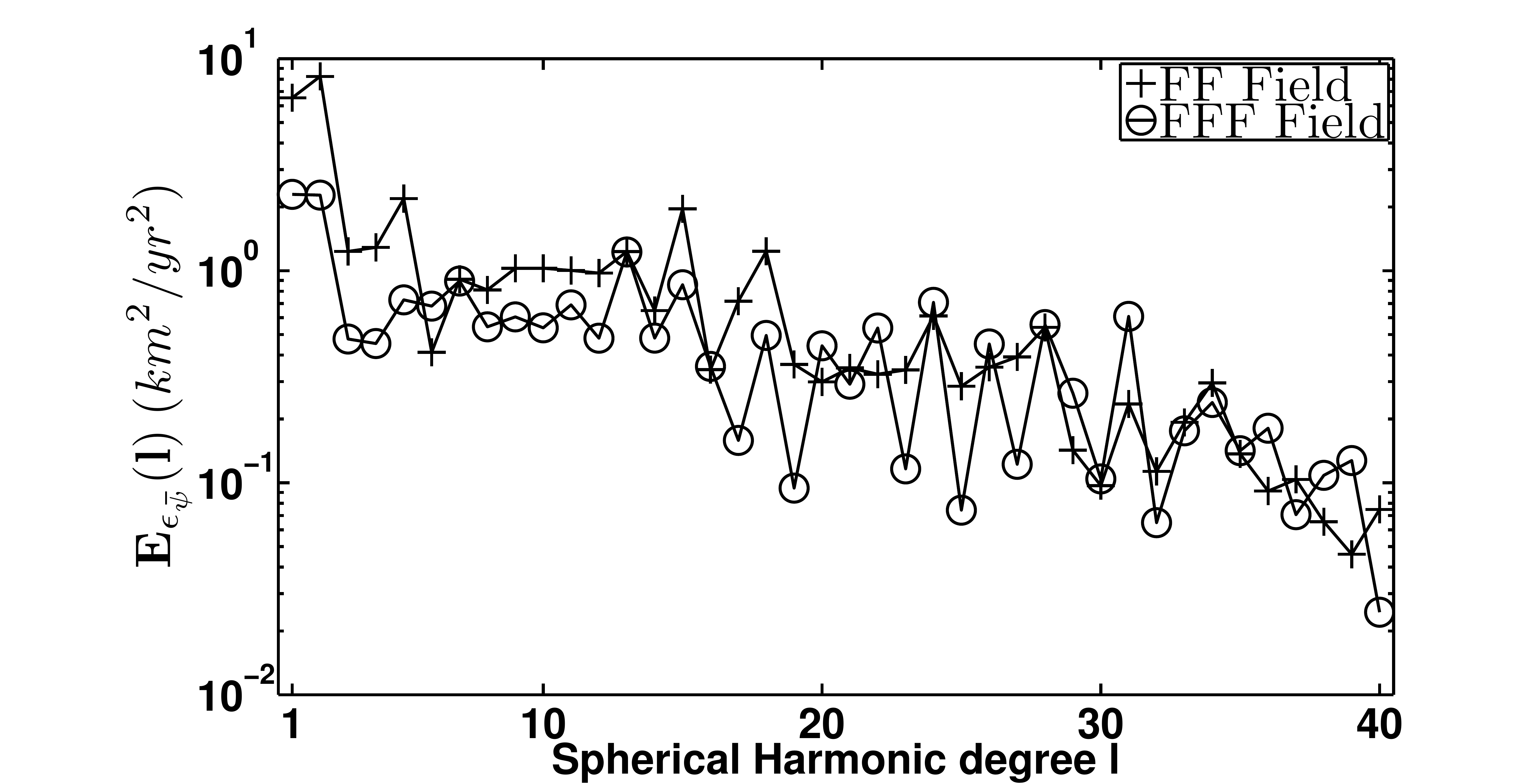}
 \end{minipage}
\caption{Poloidal (top) and toroidal (bottom) energy spectra (left) and error spectra (right). 
The full line is associated with the exact field, the plus symbols correspond to the field
solution of the FF inversion, and the circles are assigned to the field solution of the FFF
inversion.}\label{artificialSpectra}
\end{center}
\end{figure}

\noindent On the left side of figure \ref{artificialSpectra} the poloidal and toroidal 
spectra respectively defined by: 
\begin{eqnarray}
E_\Phi & = & l(l+1) \sum_{m=-l}^{m=l} (\Phi_{l,m})^2 \\
E_\psi & = & l(l+1) \sum_{m=-l}^{m=l} (\psi_{l,m})^2 \ ,
\end{eqnarray}
are plotted for the three velocity fields.
The behavior of the spectra associated with the artificial field (full line) is correctly
reproduced at large scale ($1\leq l \leq 25$) by both the FF (crosses) and 
the FFF (circles) flow models, whereas at SH degree close to the cutoff 
$l_c=40$, the discrepancies between the exact spectra and the spectra of 
the inverted fields become larger. Nevertheless, the spectra of the difference between exact
and inverted fields, displayed on the right side of figure  \ref{artificialSpectra}, show
that the use of the FFF equation allows to reproduce more accurately the artificial field
at almost any scale.


\noindent To quantify the spatial error of the two inverted fields, the following quantities were computed:
\begin{eqnarray}
\mathcal{E}_{FF} &=& \frac{\int | u  -  u_{_{FF}}|^2 \mathrm{d}\Omega}
 {\int \mathrm{d}\Omega} = 4,92 \; \text{km}^2.\text{yr}^{-2}\\ 
\mathcal{E}_{FFF} &=& \frac{\int |u - u_{_{FFF}}|^2 \mathrm{d}\Omega}
{\int \mathrm{d}\Omega} = 2,83 \;\text{km}^2.\text{yr}^{-2}\\ 
E_{tot} &=& \frac{\int  | u |^2 \mathrm{d}\Omega}
{\int \mathrm{d}\Omega} = 130 \; \text{km}^2.\text{yr}^{-2} \ ,
\end{eqnarray}
where the integration domain is the surface of the CMB, $u$ is the
exact filtered velocity field, $  u_{FF}$ and
$  u_{FFF}$ are respectively the filtered velocity field obtained by
inverting the FF and the FFF equations. The result of these computations shows
that although the energy associated to the error fields is weak in comparison
to the total energy of the exact flow, performing an inversion of FFF approximation
reduces the global error on the VF. 

%
%

\subsection{Sampling of the velocity posterior distribution}\label{sampling}
In this part we present a method to sample the posterior distribution $p(u|\gamma)$
given in equation (\ref{posteriorTot1}).
Since this distribution exhibits a complex form with respect to the velocity field $u$, directly drawing sample from it
is impossible. Nevertheless, by building an appropriate Markov Chain on the VF
one can map the distribution (for a complete description of Markov Chain Monte Carlo
methods see \cite{Gamerman2006}). For this study we chose an algorithm of the
Metropolis-Hastings type to construct the chain. The principle of the method is the following:
\begin{enumerate}[(a)]
\item in the initial step, a VF $u^n$ with $n=0$ is generated.
No particular property have to be imposed on this field, but
choosing a field which is as close as possible to the one maximizing the target distribution
will allow the chain to converge faster.
\item from $u^n$ a field $u^{n+1}$ is constructed according to some arbitrary transition kernel 
$q(u^{n+1},u^n)$.
\item the next step consists in accepting or rejecting the move from $u^n$ to $u^{n+1}$.
Therefore, an acceptance probability $\alpha(u^{n+1},u^n)$ is defined. In the case
of the Metropolis-Hastings algorithm this probability is expressed as:
\begin{equation}
\alpha(u^{n+1},u^n) = \min\left\{1, 
\frac{p(u^{n+1}|\gamma)q(u^{n+1},u^n)}{p(u^{n}|\gamma)q(u^n,u^{n+1})} \right\} \label{acceptance}\ .
\end{equation}
\item the process returns then to step (b) with $u^{n+1}$ if the move from $u^n$ to $u^{n+1}$
is accepted, and with $u^n$ otherwise.
\end{enumerate}
\noindent The ensemble is then assumed to be representative of the posterior distribution, once its averaged
field as converged towards a fixed vector.

\noindent Note that this algorithm has already been employed by \cite{Mosegaard1999,Rygaard2000} 
for sampling the posterior distribution given in equation  (\ref{posteriorTotJ}), but with a different parametrization
of the magnetic field and the velocity field. In particular, only the uncertainties on the large scale MF were considered
in these two studies.

\noindent As already mentioned previously, analytically calculating the maximum of the posterior distribution
$p(u|\gamma)$ is not feasible. We therefore decided to approximate it by taking the averaged VF of the ensemble
generated by the Markov Chain, such as:
\begin{equation}
\argmax_u p(u|\gamma) \sim \int u \; p(u|\gamma)  \mathrm{d}u\ .
\end{equation}

\subsubsection{Evaluation and comparison of the method with artificial data}\label{MCMCevaluation}
\begin{figure}[h!]
\begin{center}
\includegraphics[width=\linewidth]{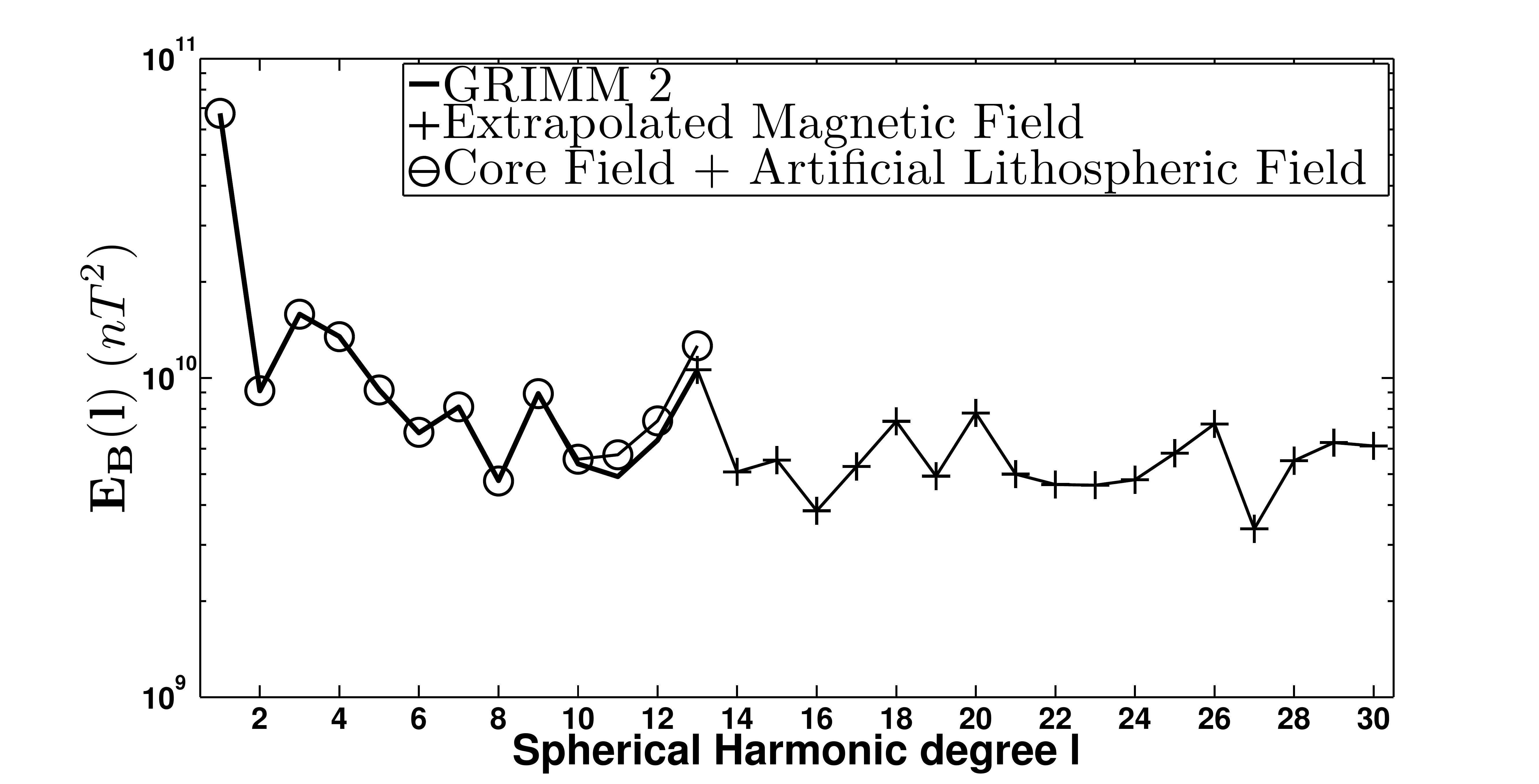}
\includegraphics[width=\linewidth]{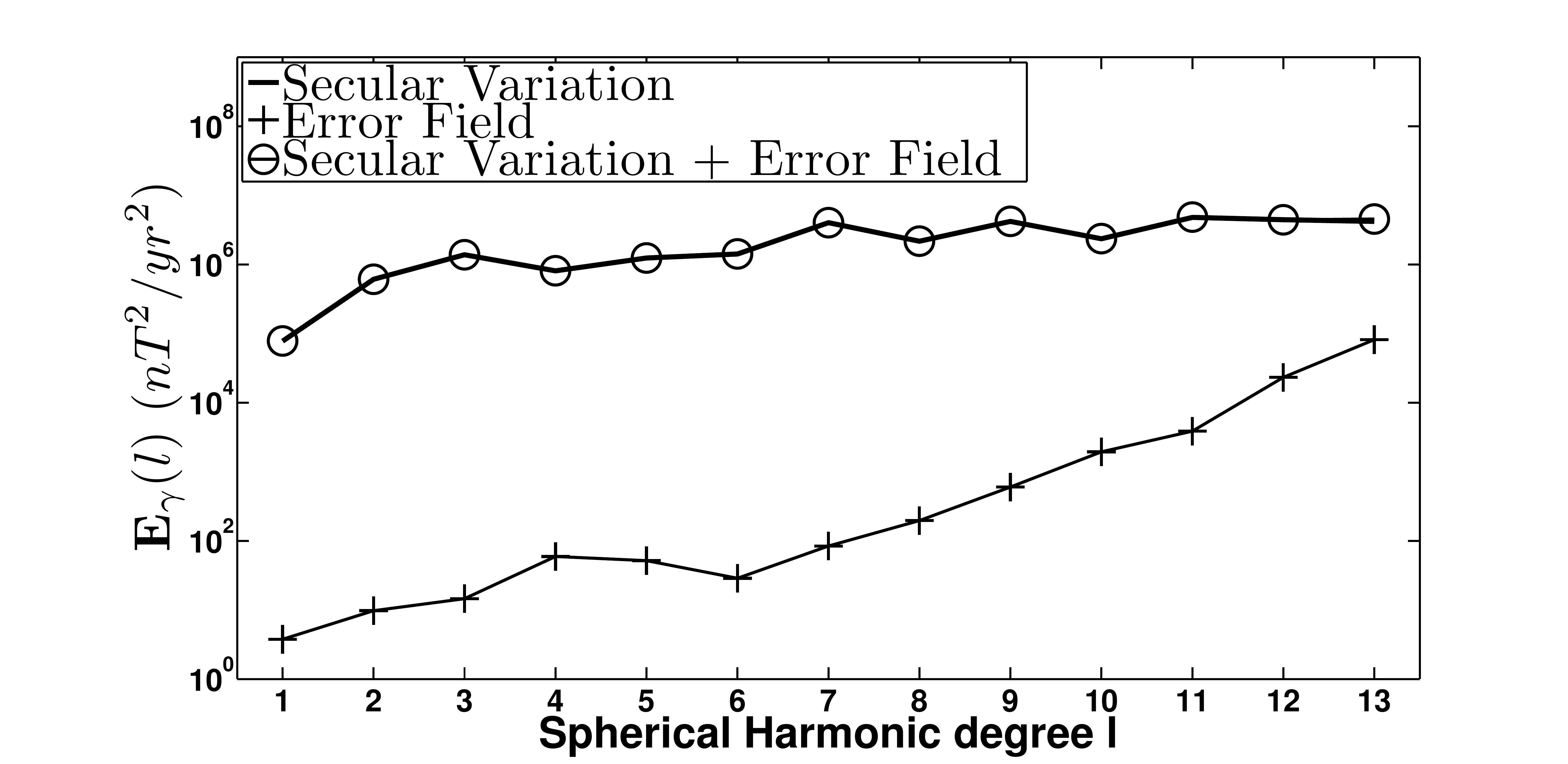}
\caption{Top: Magnetic field energy spectra. GRIMM 2 MF for the epoch $2004.0$ (thick line),
extrapolated MF (plus symbols) and GRIMM 2 MF contaminated with a randomly generated litospheric field
(circles). Bottom: Secular variation energy spectra. Artificially generated SV (thick line), error
field (plus symbols) and combination of the two fields (circles).}
\label{artificialSpectra2004}
\end{center}
\end{figure}

To evaluate our method, we created an artificial set of data, 
and performed the inversion of the FFF equation using these data.
We also compared our results to the ones obtained with alternative approaches.
The construction of the synthetic fields was performed as following:
\begin{itemize}
\item[] \textbf{Artificial magnetic field}

\noindent The large scale MF (between spherical harmonic degree $1\leq l \leq 13$) was taken from the GRIMM 2 model
at the epoch $2004.0$. Its spectrum (the thick black line in the top of figure \ref{artificialSpectra2004}) 
was then extrapolated up to degree $l=30$ according to the formulation
(\ref{buffett}) of \cite{Buffett2007}. From this small scale spectrum, and under the assumption of isotropy
and  $0$ mean of the field, a MF was randomly generated and added to the GRIMM 2 large scale field.
In figure \ref{artificialSpectra2004} (top) the spectrum of the small scale MF is represented by the plus symbols.

\item[] \textbf{Artificial velocity field}

The coefficients, in spectral space, of the poloidal ($\Phi$) and toroidal ($\psi$) fields were assumed to be 
isotropically distributed with a $0$ mean and a covariance $\tilde{\Sigma}_u$ derived from the following power law spectrum:
\begin{equation}
E_{\Phi}(l) = E_{\psi}(l) = A^2 l^{-5/3} \label{powerLawArtificial}\ ,
\end{equation}
where the value of the amplitude $A$ was chosen such as the averaged velocity intensity at the CMB was of $17$ km.yr${}^{-1}$.
A velocity field extending up to degree $l=26$ was then randomly drawn accordingly to these statistical properties.

\item[] \textbf{Artificial secular variation}

To generate an artificial large scale SV (extending up to degree $l=13$), the MF was advected by the VF through
the non-linear term of the Frozen Flux approximation (\ref{frozenflux}). The energy spectrum of the resulting SV
is presented with the thick line in figure \ref{artificialSpectra2004} (bottom).

\end{itemize}

\noindent The next step of this evaluation was to recover the velocity field according to the magnetic
field and the secular variation. But for this inverse problem to be more realistic, uncertainties were added to the data.
A large scale lithospheric field ($1\leq l \leq 13$) was randomly generated accordingly to the theoretical spectrum
$E_B^L$ of \cite{Thebault2013} (see equation \ref{litosphericSpectrum}) and added to the artificial MF.
This contaminated field was then truncated at degree $l=13$. It is referred as $b_0$ and its energy spectrum is plotted 
on top of figure \ref{artificialSpectra2004} with circles. The uncertainties on the secular variation
were built by randomly drawing a field from a Gaussian distribution with a $0$ mean and a covariance given
by the posterior covariance matrix of the GRIMM 2 secular variation $\Sigma_\gamma$. The resulting field was then superimposed on
the artificial field. The spectrum associated with the total SV is shown with circles on the bottom of figure \ref{artificialSpectra2004}.
\vspace{0.5cm}

\noindent Except for the prior distribution of the velocity field $p(u)$, all the distributions derived in section \ref{baysian}
are consistent to describe the posterior distribution of the velocity field in this test. We recall that they read:
\begin{eqnarray}
p(\gamma|u,b) & = & \mathcal{N}\left( -\left({\bf \nabla_{_H} }\left(ub+\tau\right)\right)^<,\Sigma_\gamma\right) \ ,\label{likelyTest} \\
p(b) & = & \mathcal{N}( b_0,\Sigma_b) \label{priorBTest} \ ,
\end{eqnarray}
where $\mathcal{N}(x_0,\Sigma_x)$ corresponds to the normal distribution centered in $x$ and with covariance $\Sigma_x$,
and ${\Sigma}_b$ is the covariance of the MF in which are included the uncertainties due to the lithospheric field
and the modelization of the unresolved MF.
For this evaluation phase, the prior distribution of the VF was given by:
\begin{equation}
p(u)  =  \mathcal{N}( 0,\tilde{\Sigma}_u) \ ,\label{priorUTest} 
\end{equation}
with $\tilde{\Sigma}_u$ the covariance of the VF derived from the power laws (\ref{powerLawArtificial}).

\noindent The estimation of the flow with respect to the artificial MF and SV was then realized with the four algorithms 
presented below.

\begin{itemize}

\item[] \textbf{MCMC method}

\noindent In this approach the full posterior distribution of the velocity 
field $p(u|\gamma)$ is explored with the Metropolis-Hastings algorithm described in the beginning of the section. 
The transition kernel $q(u^{n+1},u^n)$ entering the algorithm (step b)
is derived from the prior distribution of the VF as following:
\begin{equation}
q(u^{n+1},u^n) = \frac{exp\left[{-\frac{1}{2\lambda^2} (u^{n+1}-u^n)^T 
\tilde{\Sigma}_u^{-1} (u^{n+1}-u^n)}\right]}{(2\pi)^{\frac{d}{2}} |\tilde{\Sigma}_u|^\frac{1}{2}} 
\label{transition}\ ,
\end{equation} 
where the factor $\lambda$ allows to rescale the covariance $\tilde{\Sigma}_u$ in order
to limit the distance of the move from one field to the other. Since this kernel
is symmetric with respect to $u^{n+1}$ and $u^n$, the acceptance probability
$\alpha(u^{n+1},u^n)$ of equation (\ref{acceptance}) reduces to:
\begin{equation}
\alpha(u^{n+1},u^n) = \min\left\{1,\frac{p(u^{n+1}|\gamma)}{p(u^{n}|\gamma)} \right\} \ .
\end{equation}
The initial field was set to $0$ and an ensemble of $N_m = 230000$ VF $u_i$ was generated.
We then approximated the flow maximizing the posterior distribution through the averaging operation:
\begin{equation}
u_0  =  \frac{1}{N_m} \sum_{i=1}^{i=N_m} u_i \ .\label{priorUTest} 
\end{equation}

\item[] \textbf{Least square method}

\noindent In this method, the magnetic field $b$ is assumed to be exactly known such as:
\begin{equation}
p(b)  =  \delta(b-b_0) \ .\label{priorUTest} 
\end{equation}
As a consequence, the posterior distribution of the velocity field becomes proportional to:
\begin{eqnarray}
 p(u|\gamma)  &\sim& 
\exp\left[-\frac{1}{2} 
\left(\gamma + A_{b_0 }u \right)^T 
\Sigma_{{\gamma}}^{-1}
\left(\gamma + A_{b_0  }u \right)\right]  \nonumber \\
 & \times & \exp\left[-\frac{1}{2} 
 u^T \tilde{\Sigma}_u^{-1} u \right] \label{leastSquare}\ ,
\end{eqnarray}
where $A_{b_0}$ is the operator allowing to evaluate the non linear term of the FFF equation when
it is applied to $u$. The maximum of this distribution can be calculated analytically and corresponds to the usual
least square solution:
\begin{equation}
u_0 = -(A_{b_0 }^T\Sigma_{{\gamma}}^{-1}A_{b_0 }+\tilde{\Sigma}_u^{-1})^{-1} A_{b_0}^T\Sigma_{{\gamma}}^{-1}\gamma \label{leastSquareSolution}\ .
\end{equation}

\item[] \textbf{Ensemble method}

\noindent This approach was developed by \cite{Gillet2009} and consists in generating an ensemble of magnetic field
and calculating their associated velocity field. In this test,
the ensemble of MF was randomly drawn from the distribution $p(b)$ given in equation (\ref{priorBTest}).
In total we generated $N_e=100$ magnetic fields ${b_i}$ (with $1..i..N_e$) extending up to degree $l=30$. 
Each velocity field $u_i$ were then determined by the relation:
\begin{equation}
u_i = -(A_{b_i}^T\Sigma_{{\gamma}}^{-1}A_{b_i }+\tilde{\Sigma}_u^{-1})^{-1} A_{b_i}^T\Sigma_{{\gamma}}^{-1}\gamma \ .
\end{equation}
where the operator $A_{b}$ now depends on each realization $b_i$ of the magnetic field.
The flow solution of the complete inverse problem is given by:
\begin{equation}
u_0 = \frac{1}{N_e} \sum_{i=1}^{i=N_e} u_i \ .
\end{equation}

\item[] \textbf{Iterative method}

\noindent In this approach, \cite{Lesur2013} proposed to determine the CMB velocity field through the following
iterative process:
\begin{equation}
u^{n+1} = -(A_{b_0}^T\left(\Sigma_{\tilde{\gamma}}^n\right)^{-1}A_{b_0}+\tilde{\Sigma}_u^{-1})^{-1} A_{b_0}^T\left(\Sigma_{\tilde{\gamma}}^n\right)^{-1}\gamma \label{iterativeSolution}
\end{equation}
where $n$ is the index of the iteration, and the covariance $\Sigma_{\tilde{\gamma}}^n$ is given by:
\begin{equation}
\Sigma_{\tilde{\gamma}}^n =  \Sigma_{\gamma} + \left(A_{u^n}\right) \Sigma_b \left(A_{u^n}\right)^T \label{iterativeLesur}
\end{equation}
with $A_{u^n}$ the operator depending on the velocity field $u^n$, and which allows to evaluate the non linear term of the FFF equation
when it is applied to $b$. In the numerical implementation of this method, the initial field $u^0$ was taken as the one 
solution of the least square approach.

\noindent Note that this algorithm provides an estimation of the maximum of the posterior distribution given in equation (\ref{posteriorTot1})
if the quantity $A_{u} \Sigma_b A_{u}$ is assumed to vary slowly with respect to $u$.
\end{itemize}
 \vspace{0.5cm}
 
\noindent In the previous section we mentioned the algorithm of \cite{Pais2008} which also allows to take into account the
variations of the MF in the inverse problem. We decided not to implement it in this evaluation since it is an approximation
of the method proposed by \cite{Lesur2013}.

\vspace{0.5cm}

\noindent To compare the different approaches, the artificial velocity field $u$ is decomposed into
a toroidal ($\psi$) and a poloidal ($\Phi$) field, with $\psi_{l,m}$ and $\Phi_{l,m}$ their 
respective spectral counterpart. The same operation is performed 
on the velocity fields $u_0$ given by the four inversion methods. Their toroidal and poloidal
parts are then referred as $\psi_0$ and $\Phi_0$ in physical space, and $\psi_{l,m}^0$ 
and $\Phi_{l,m}^0$ in spectral space.

\noindent To measure the accuracy, scale by scale, of the various fields, we define the poloidal
and toroidal error fields as:
\begin{eqnarray}
\epsilon_{l,m}^\Phi & = & \Phi_{l,m} - \Phi_{l,m}^0 \label{poloError2004}\\
\epsilon_{l,m}^\psi & = & \psi_{l,m} - \psi_{l,m}^0 \label{toroError2004}\ ,
\end{eqnarray}
and their associated energy spectra:
\begin{eqnarray}
E_{\epsilon_\Phi} & = & l(l+1) \sum_{m=-l}^{m=l} (\epsilon_{l,m}^\Phi)^2 \\
E_{\epsilon_\psi} & = & l(l+1) \sum_{m=-l}^{m=l} (\epsilon_{l,m}^\psi)^2 \ .
\end{eqnarray}

\begin{figure}[h!]
\begin{center}
\includegraphics[width=\linewidth]{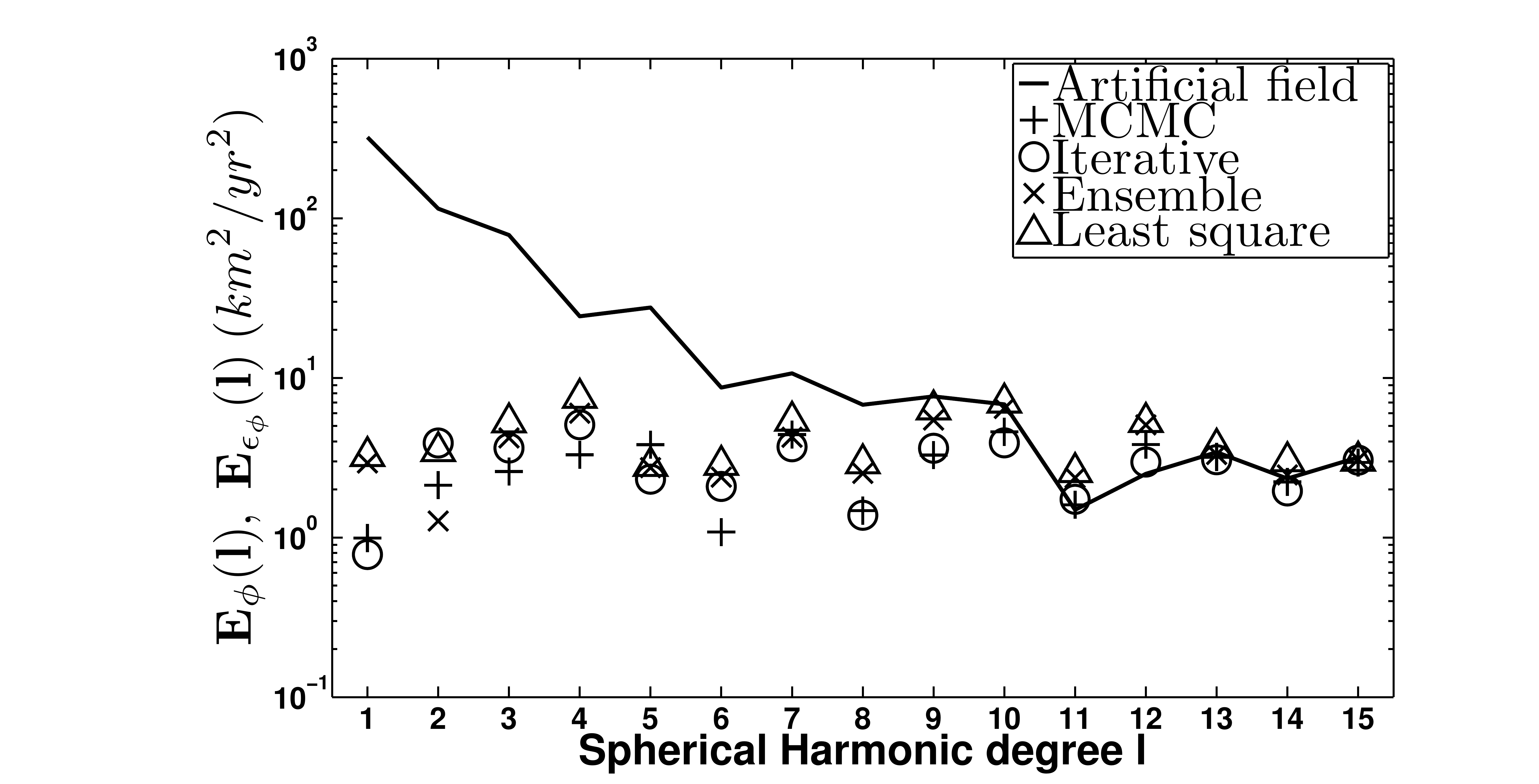}
\includegraphics[width=\linewidth]{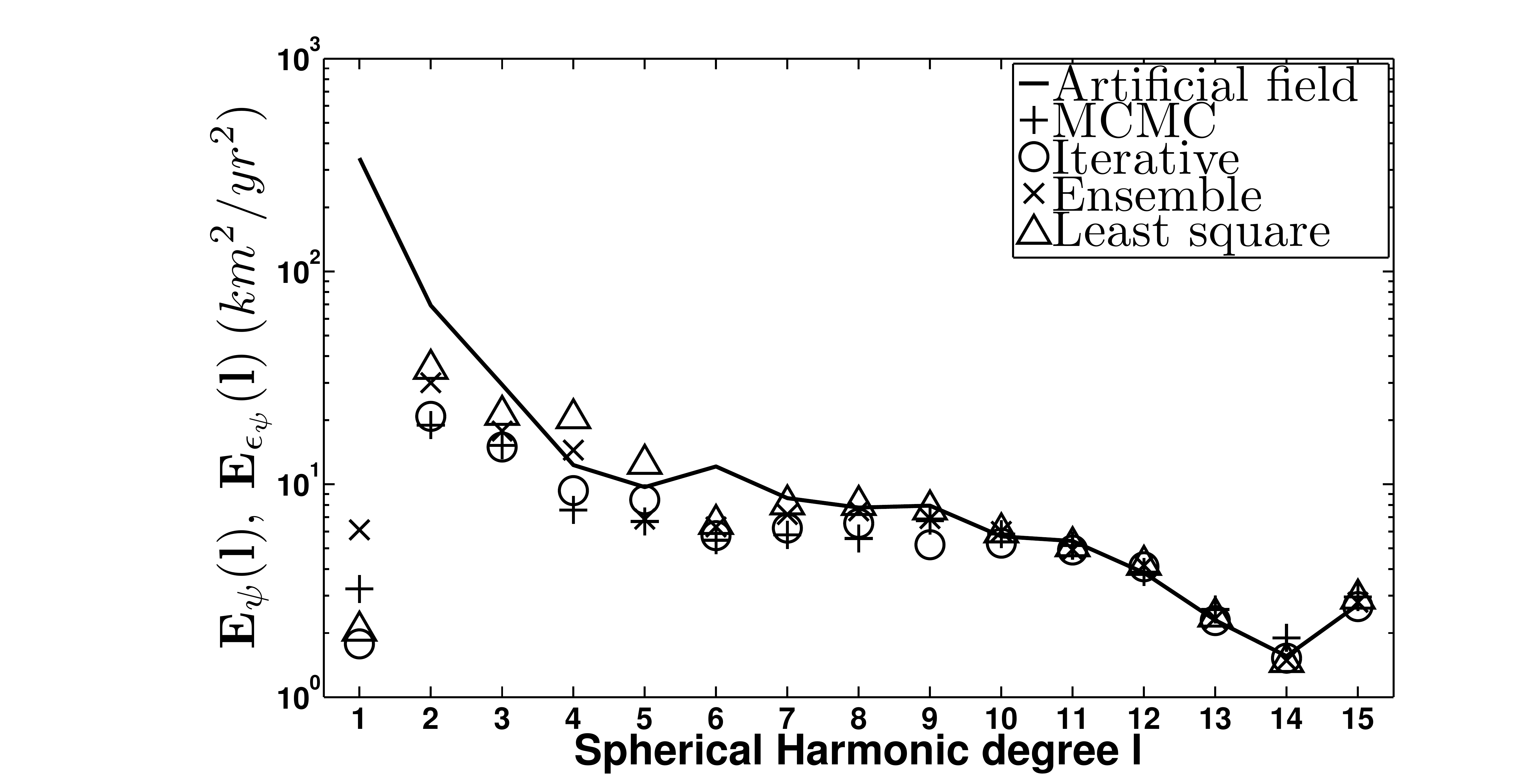}
\caption{Poloidal (top) and toroidal (bottom) energy spectra for the artificial
VF (thick lines) and for the different error fields (symbols).}
\label{errorSpectra2004}
\end{center}
\end{figure}

In figure \ref{errorSpectra2004} the spectra of the error fields (symbols) are plotted together
with the poloidal and toroidal spectra of the artificial velocity field (solid lines). 
The first observation one can make is that above the SH degree $l=10$, the energy 
associated with the poloidal and toroidal error is of the same order than the energy 
of the artificial fields. This implies that above this degree, the estimation of the flow
is not reliable whatever the method employed to determine it. At large scale, the performance
of the different approaches varies strongly. Whereas the flow obtained with the least square method 
(triangles) is the one which deviates the most from the artificial flow, the velocity fields evaluated with 
the iterative method (circles) and the MCMC algorithm (plus symbols) are the ones presenting
the lowest error intensities. In between, in terms of accuracy, lies the ensemble
method (crosses). It can be noted that at degree $l=5$ and between $l=5$ and $l=6$ for respectively
the ensemble and the least square method, the energy associated with the toroidal error field is larger
than the energy of the artificial field itself. This is not the case anymore when using the iterative or MCMC
approaches.

\begin{figure}[h!]
\begin{center}
   \includegraphics[width=\linewidth]{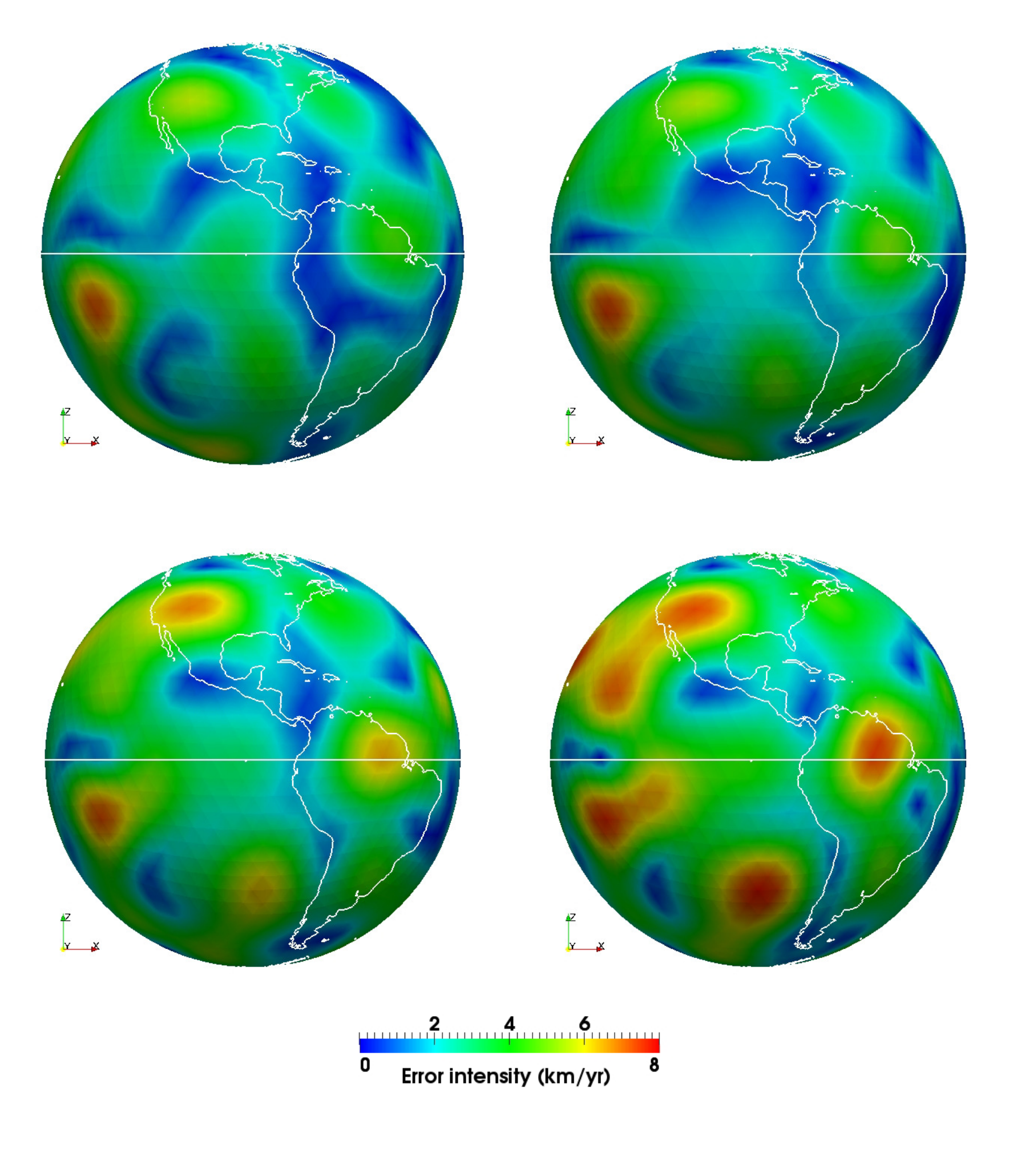}
\caption{Intensity, in km.yr${}^{-1}$, of the difference between the artificial velocity field and the velocity fields evaluated
with the following approaches: MCMC (top left), iterative from \cite{Lesur2013} (top right) ensemble from \cite{Gillet2009} 
(bottom left) and least square (bottom right). }\label{errorField2004}
\end{center}
\end{figure}

\noindent To evaluate the different models in physical space, the coefficients associated with the poloidal and toroidal error fields
(equations \ref{poloError2004}-\ref{toroError2004}) were truncated at degree $l=10$, since at smallest scales the uncertainties
are maximal, and projected in real space. Figure \ref{errorField2004} shows the intensity of these fields 
at the level of America for the four approaches. Although the locations of the errors are similar, 
their intensities differs strongly from one flow to the other. A computation of the averaged energy associated with the poloidal and toroidal
error fields (see table \ref{tableError}) shows that globally the best approximation of the artificial velocity field
is provided by the MCMC algorithm, followed in order by the iterative, ensemble and least square methods.
Note that the differences between the MCMC and iterative approach are very low, and since the computation time required
to sample the full posterior distribution is much larger than the one to approximate its maximum 
using the iterative method, if one wants to determine the flow without its underlying uncertainties, using the algorithm of
\cite{Lesur2013} is certainly more appropriate.

\begin{table}
\caption{Averaged energy of the error in km${}^2.$yr${}^{-2}$ associated with the different inverted velocity fields.
These values have to be compared with the averaged energies of the artificial poloidal and toroidal fields
which are respectively of $48.07$ km${}^2.$yr${}^{-2}$ and $45.28$ km${}^2.$yr${}^{-2}$.}
\centering
\begin{tabular}{l c c c c}
\hline
  & MCMC & Iterative & Ensemble & Least square   \\
\hline
 Poloidal Field  & $2.20$ & $2.42$ & $3.04$ & $3.72$ \\
\hline
 Toroidal Field  & $6.45$ &  $6.72$  &  $8.69$ & $10.1$ \\
\hline
\label{tableError}
\end{tabular}
\end{table}

\begin{figure}[h!]
\begin{center}
   \includegraphics[width=\linewidth]{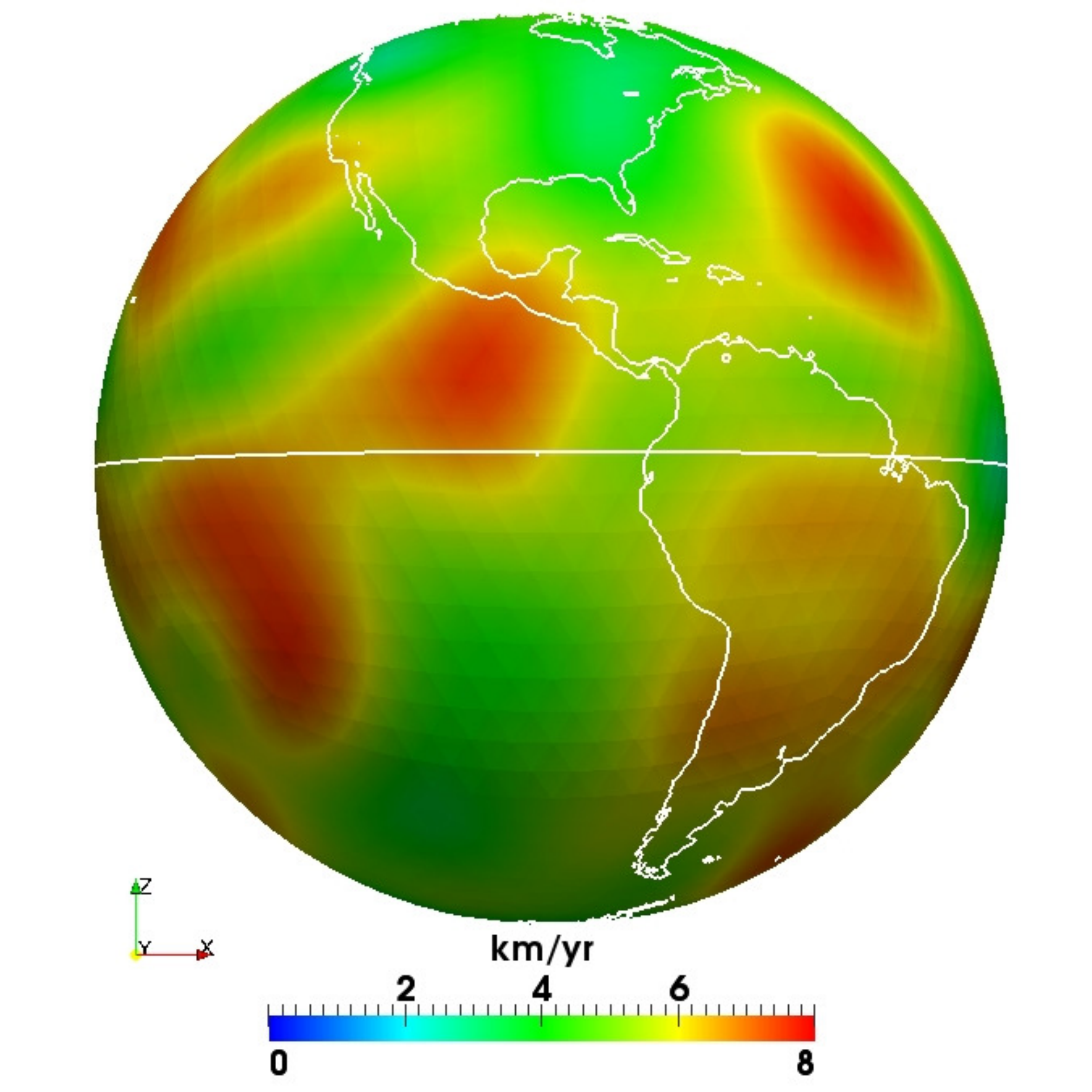}
\caption{$95\%$ confidence interval on the velocity intensity in km.yr${}^{-1}$ according to the MCMC flow ensemble.}\label{SDevaluation}
\end{center}
\end{figure}

\noindent Since the MCMC algorithm provides an information on the flow uncertainties, we extracted the standard
deviation of the velocity intensity $\sigma{|u|}$, from the variance $\sigma_u^2$ as following:
\begin{equation}
\sigma_u^2  =  \int diag\left[\left(u-u_0\right)\left(u-u_0 \right)^T\right]
p(u|\gamma) \mathrm{d}u 
\end{equation}
where the term $diag$ means that only the diagonal elements of the matrix lying within
the brackets are kept. At each node of the discrete CMB, the VF $u$ is composed of a polar and an azimuthal component, so
to get the variance associated with the velocity intensity $\sigma_{|u|}^2$, the variance of each component has to be summed up. 
In figure \ref{SDevaluation} the quantity $2\sigma_{|u|}$, corresponding to the $95\%$ confidence interval on the flow intensity,
is displayed at the level of $America$. This picture presents a pessimistic view of the uncertainties to be expected when evaluating the flow
maximizing the posterior distribution. Because the probability for the real flow to lie within the tails of the posterior
distribution is very low, the predicted error will globally be larger than the effective one.
Nevertheless, locations where the differences between the solution flow and the real one are important corresponds to
areas of high posterior variance.  

\noindent Finally, we wanted to measure the impact of the velocity field prior information on the solution
of the inverse problem. We therefore realized a simulation with the iterative algorithm of \cite{Lesur2013},
in which only the covariance matrix of the velocity field $\tilde{\Sigma}_u$ of equation (\ref{iterativeLesur}) has been
modified. Instead of being the covariance imposed to generate the artificial velocity field,
this latter was replaced by the covariance ${\Sigma_u}$ defined in equation (\ref{norm}) and derived from the Bloxham's "strong norm".
We then calculated, as previously, the averaged energy of the error field using the velocity field we obtained.
It is of $3.45$ km${}^2.$yr${}^{-2}$ and $9.23$ km${}^2.$yr${}^{-2}$ for respectively the poloidal and toroidal part of the error.
This shows us that the choice of the prior information on the velocity field influences in a significant manner 
the results of the inverse problem. Nevertheless, it can be noticed that the intensity of the error in this case
remains lower than the intensity of the error for the flow resulting from the least square approach, showing
the importance of modeling the possible variations of the MF.

\subsubsection{Application of the MCMC algorithm for the epoch 2005.0}
In the simulation we realized, the MF and the SV as well as the covariance for the SV,
are given up to SH degree $l=13$ by the GRIMM 2 model of \cite{Lesur2010} for the epoch $2005.0$.
For the magnetic field, the covariance associated with its large scales ($0<l\leq 13$) is derived
from the theoretical spectrum of the litospheric field of \cite{Thebault2013} as shown in equations 
(\ref{litosphericSpectrum})-(\ref{LSMFcovariance}), whereas the extrapolation of the MF spectrum 
proposed by \cite{Buffett2007} is used to build the covariance matrix of the small 
scale MF ($13<l\leq 30$) as presented in equation (\ref{SSMFcovariance}).
The velocity field is extended up to SH degree $l=26$, and its prior distribution is given
by the relation (\ref{priorU}). The width of the filter has been set to $\overline{\Delta}=500$km, and
the initial field of the Markov chain $u_0$ , is the velocity field solution of the least square approach:
\begin{equation}
u_0 = -\left( A_B^T \Sigma_\gamma^{-1} A_B +  \Sigma_u^{-1} \right)^{-1}A_B^T\Sigma_\gamma^{-1}
\gamma \label{u0}
\end{equation} 
with $A_B u = \left({\bf \nabla_{_H} }\left(ub+\tau\right)\right)^<$. 

\noindent As for the synthetic test of section \ref{MCMCevaluation}, the transition kernel $q(u^{n+1},u^n)$ was 
derived from the prior distribution of the VF as shown in equation (\ref{transition}).

\noindent The Metropolis-Hastings algorithm was then numerically simulated on a discrete CMB refined 4 times.
An ensemble of $130000$ VF $u$ mapping the posterior distribution have been generated. The velocity field  $\hat{u}$ 
maximizing the posterior, is approximated by taking the average field of the ensemble we created.

\begin{figure}[h!]
\begin{center}
     \includegraphics[width=\linewidth]{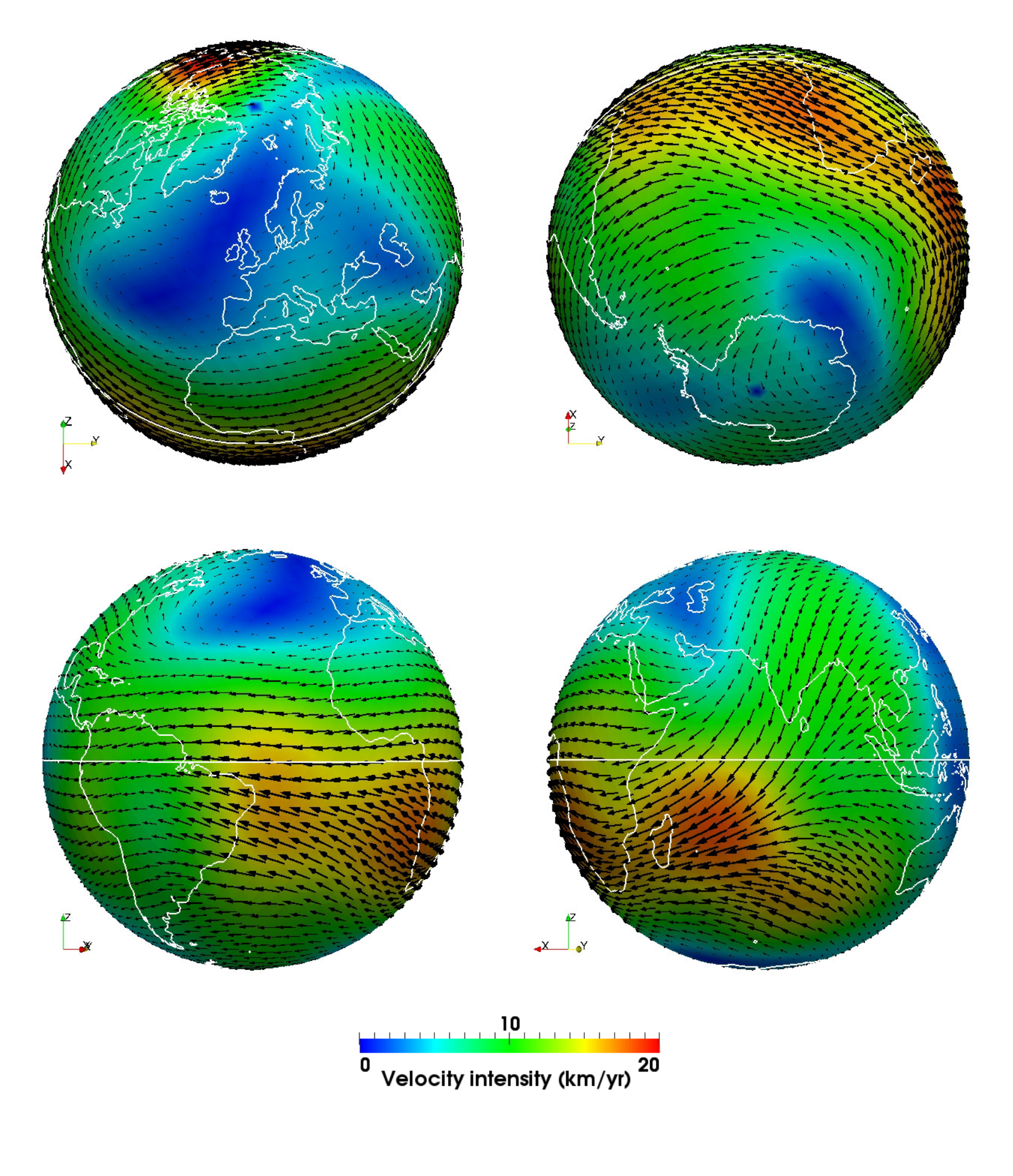}
\caption{Velocity field $\hat{u}$ and its intensity for the epoch $2005.0$.}
\label{VF2005}
\end{center}
\end{figure}

\noindent In figure \ref{VF2005}, the vector field $\hat{u}$ and its intensity
are displayed in different locations of the core mantle boundary. Many features
of the flow we obtained have already been reported in previous studies.
In particular, the eccentric and planetary scale anticyclonic gyre observed by \cite{Pais2008} and \cite{Gillet2009}
is also present in the flow we calculated. This observation reinforces the hypothesis that the fluid motions 
in the outer core can be well-described by the compressible quasi-geostrophic assumption, a constraint
a priori applied in both studies.
Nevertheless, according to our results, deviations from quasi-geostrophy (QG) have also to be expected. Indeed, under this
hypothesis, the flow is forced to be symmetric with respect to the equator outside the tangential cylinder, a condition which is not 
fulfilled everywhere in our case. Although the symmetric part of the velocity field is dominant in our simulation ($82\%$ of
the energy of the total VF is concentrated in its symmetric components) certain patterns, such as
the flow crossing the equator below India or the larger intensity of the westward drift in the southern hemisphere 
are violating this property.
The flow we obtained also exhibits a much smoother spatial behavior than the ones presented by
\cite{Pais2008,Gillet2009}.
This is certainly due to the choice we made to characterize a priori the velocity field. Through the Bloxham's "strong norm",
we imposed a very steep spectrum (in $l^{-5}$) to both the poloidal and toroidal field, as a consequence
the intermediate scales of the VF were probably over-damped in our simulation.

\begin{figure}[h!]
\begin{center}
     \includegraphics[width=0.9\linewidth]{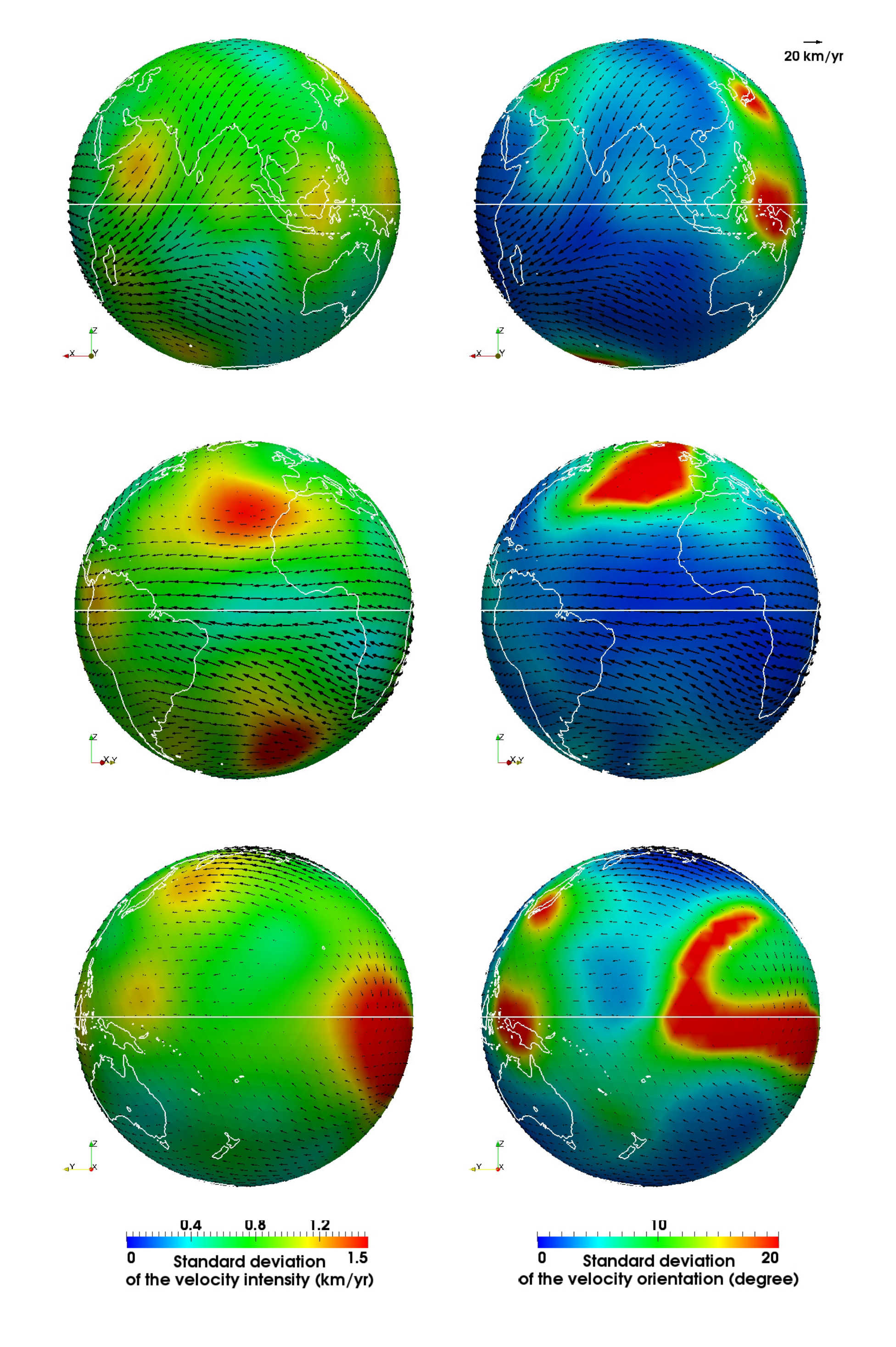}
\caption{Velocity field $\hat{u}$ for the epoch $2005.0$ and its associated uncertainties on the intensity
$\sigma_{|u|}$ in km.yr${}^{-1}$ (left) and orientation $\sigma_\Gamma$ in degree (right).}
\label{uncertainties2005}
\end{center}
\end{figure}

\noindent Possessing the full posterior distribution of the VF allows to extract many useful information
on the flow. In particular the uncertainties on the velocity field intensity and orientation
(with respect of course to the prescribed modelization)
can be evaluated at any spatial location on the grid. To compute the standard deviation of the velocity
field intensity $\sigma_{|u|}$ we followed the protocol given in section \ref{MCMCevaluation}, whereas
the standard deviation of the velocity orientation $\sigma_\Gamma$, is derived from the formula:
\begin{equation}
\sigma_\Gamma^2  =  \int diag\left[\Gamma\Gamma^T\right] p(u|\gamma) \mathrm{d}u \ ,
\end{equation}
where $\Gamma$ is the angle, in degree, between the velocity field $u$ and $\hat{u}$. 
The quantities $\sigma_{|u|}$  and $\sigma_\Gamma$, 
are displayed in figure \ref{uncertainties2005} (on the left for the $\sigma_{|u|}$
and on the right for the $\sigma_\Gamma$). This figure first shows that a strong (resp. weak) uncertainty
on the intensity coincides with a strong (resp. weak) uncertainty on the orientation of the
flow. Secondly one can notice that the uncertainties are not homogeneously distributed
on the surface of the CMB. While the planetary scale eccentric gyre seems to be very robust, the VF at the 
level of the Pacific ocean is much more uncertain. It is known that in this latter area
the magnetic field activity is moderate (\cite{Hulot2002}). As a consequence the secular variation
is low, and its associated uncertainties, due to the inaccuracy of the measurements and 
to the interactions between the unresolved MF and the VF, may be larger than the signal itself.
It is therefore very difficult to evaluate the velocity field in this region. 
Nevertheless, this maps shows that there is a large
part east of Australia and around the longitude of New-Zealand  where the VF can 
be accurately estimated. Another particular feature can be noticed at the level of the
North-Atlantic ocean, where the uncertainties on both the flow intensity and orientation are very
large. As already questioned by \cite{Finlay2010}, the robustness of the clockwise gyre usually observed
in this area by models assuming Tangential-Geostrophy seems to be very weak according to our results.

\begin{figure}[h!]
\begin{center}
\includegraphics[width=\linewidth]{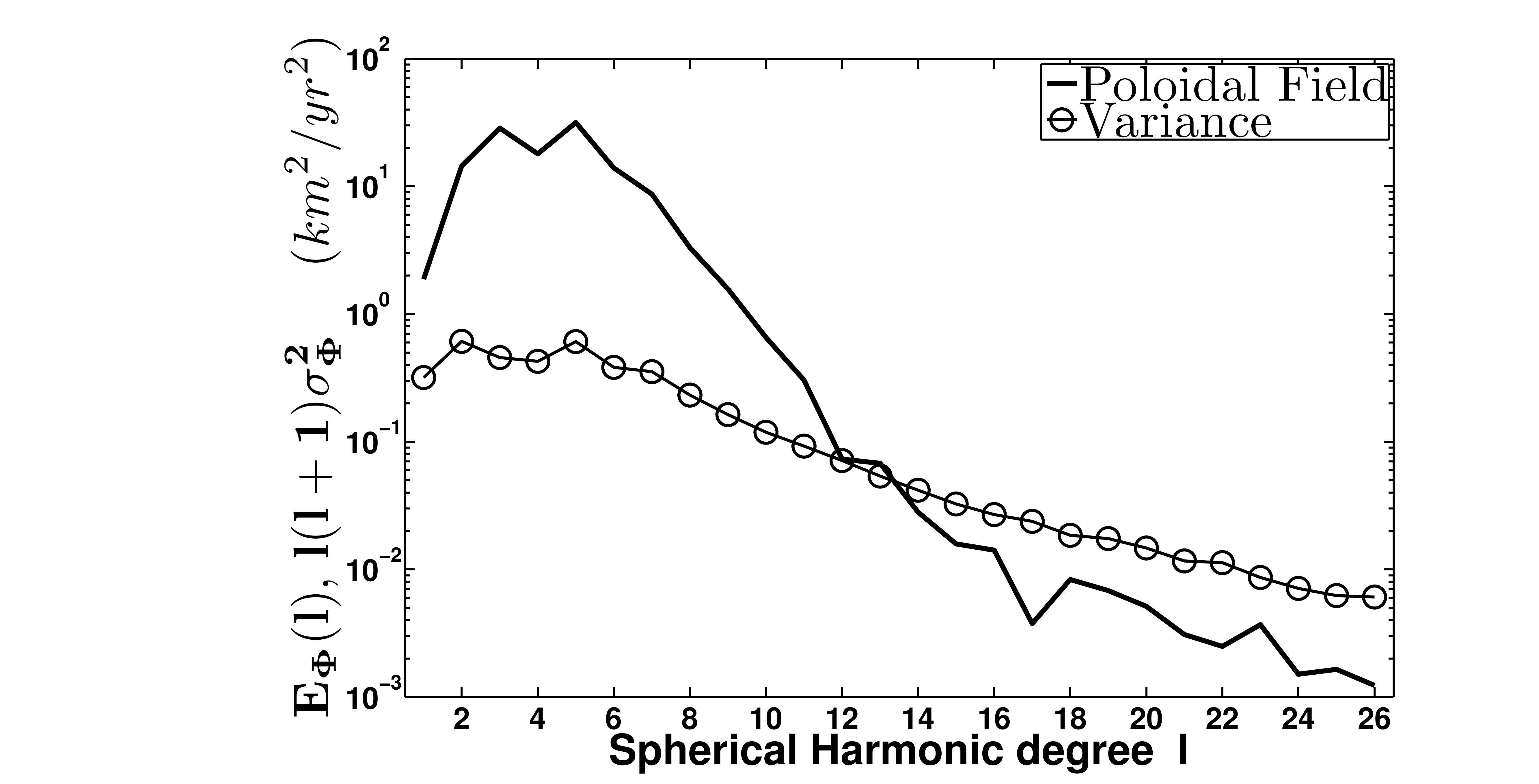}
\includegraphics[width=\linewidth]{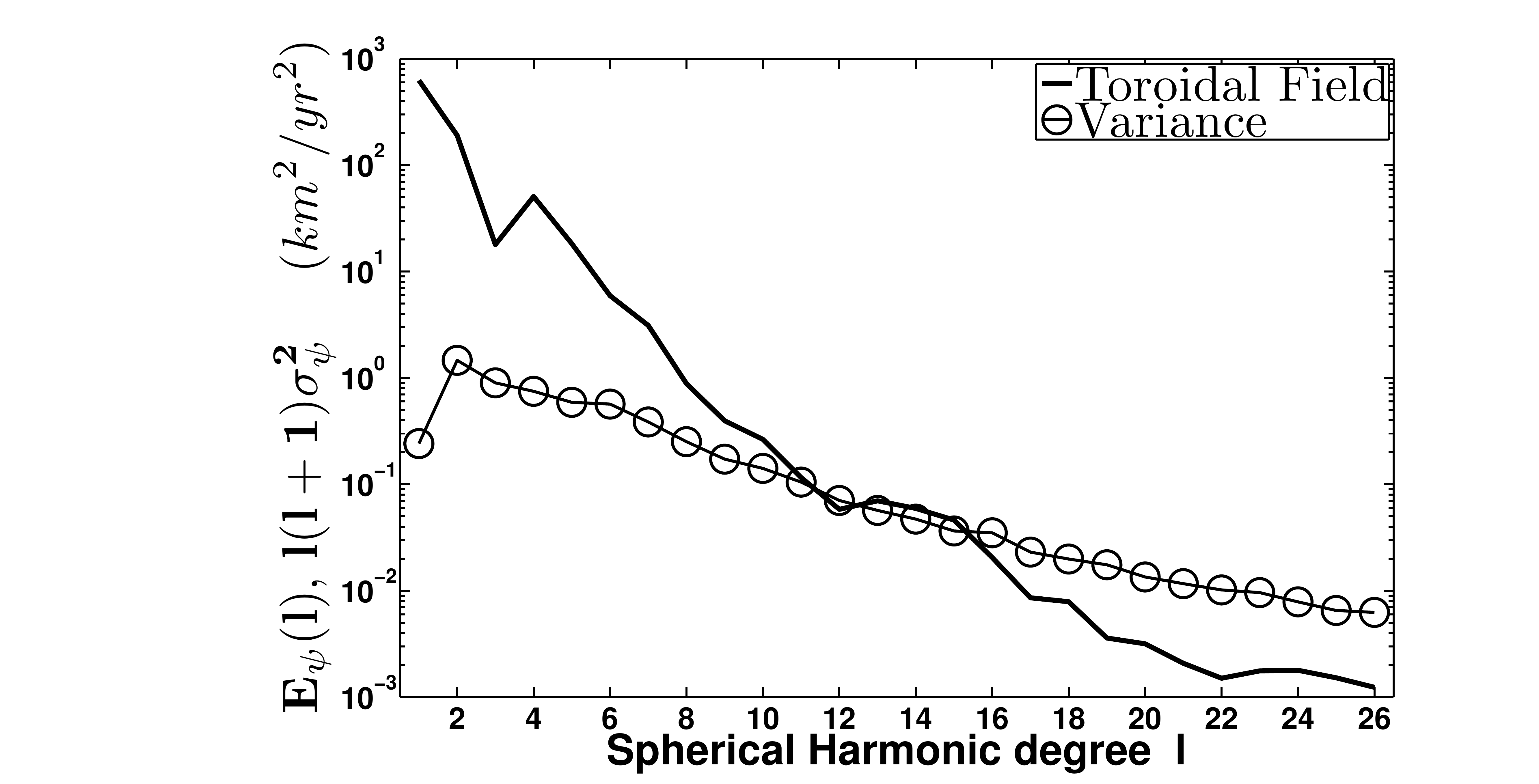}
\caption{Poloidal (top) and toroidal (bottom) energy tospectra associated with the
velocity field $\hat{u}$ (thick line),  and with the
spectral uncertainties (circles).}
\label{spectralUncertainty}
\end{center}
\end{figure}

\noindent Since we are using truncated MF and SV it may be interesting to investigate the spectral 
properties of the posterior VF. The poloidal $\Phi$ and toroidal $\psi$ part of the VF $u$ 
are expanded in spherical harmonics. The resulting fields are respectively referred as $\Phi_{l,m}$ 
and $\psi_{l,m}$. The same operation is performed on $\hat{u}={\bf{\nabla}} \hat{\Phi} + {\bf{e_r}} 
\times {\bf\nabla} \hat{\psi}$, with $\hat{\Phi}_{l,m}$ and $\hat{\psi}_{l,m}$ its poloidal and
toroidal SH coefficients.
In figure \ref{spectralUncertainty} are plotted the spectra associated with $\hat{\Phi}_{l,m}$ and $\hat{\psi}_{l,m}$
as well as the variance on these coefficient summed up over the order $m$ and rescaled by the factor $l(l+1)$.
We can observe that the largest scales of the flow are dominated by the toroidal field.
A computation of the poloidal and toroidal energy shows that more than $88\%$ of the total energy
is of toroidal nature. The other information which can be extracted from figure 
\ref{spectralUncertainty} is that, as for the synthetic test we realized previously,
above the SH degree $l=10$, the intensity of the flow uncertainties becomes larger
than the intensity of the flow itself. As a consequence the evaluation of the small scale
velocity field cannot be considered as reliable.

\noindent It has to be emphasized that all the results we obtained are conditioned by the 
choice of the prior information imposed to the flow, and should not be considered as absolute.
In particular the globally low level of uncertainties on the flow solution is certainly a consequence
of the strong regularization we employed.

\section{Conclusions}\label{conclusion}

In this study we have presented a new method to determine the velocity field at the Earth's core mantle
boundary according to an outer core magnetic field and secular variation model. We showed
that using an appropriate dynamical equation to prescribe the large scale magnetic field evolution in the inverse problem,
permitted to reduce the modeling errors arising from the truncated nature of the available fields.
We also demonstrated that the Bayesian formalism we developed to account for the large scale uncertainties
on the magnetic field and to model the unresolved small scale MF, allowed to properly
describe the inverse problem as soon as the information introduced a priori were accurate.
Through the evaluation of our method and the comparison with other approaches, we could indirectly 
confirm that the unresolved part of the magnetic field contributed significantly to the observed secular variation,
and that its modelization was necessary to obtain a more accurate description of the flow at the core mantle boundary.

\noindent When we applied our method to real data, provided by the GRIMM 2 model for the epoch $2005.0$,
we could recover many features of the flow already observed in previous studies where
a different prior information on the velocity field had been considered. In particular we could retrieve
the planetary-scale eccentric gyre characteristic of flow evaluated under the compressible
quasi-geostrophy assumption (see \cite{Pais2008,Gillet2009}). Nevertheless, according to our simulation, 
the flow is crossing the equator below India
and the intensity of the westward  drift is the larger in the southern hemisphere, indicating that the 
equatorial symmetry imposed by the quasi-geostrophy hypothesis is broken. Through this observation
one can conclude that deviations from quasi-geostrophy should be allowed to occur when
this latter constraint is imposed in the inverse problem. Another specificity of the velocity field we obtained is its 
very smooth spatial behavior. This property is certainly induced by the prior distribution of the velocity field we chose,
since this latter imposes a very steep spectrum to both the poloidal and toroidal part of 
the velocity field. 

\noindent Finally, thanks to the ensemble of velocity field we generated to map the posterior distribution, we could
evaluate the uncertainties associated with the flow solution of the inverse problem.
According to our results, whereas on the one hand, the robustness of the flow is questioned in many area, 
and particularly in almost the entire Pacific ocean, and in the northern part
of the Atlantic ocean, on the other hand, the planetary-scale eccentric gyre seems to be a very robust structure.
From the evaluation of the uncertainties in spectral space, we concluded that the flow at the CMB could only
be accurately estimated at large scales (between spherical harmonics degree $1$ and $10$).

\appendix
\section{Spherical diffusion}\label{appA}

In Cartesian space, applying a Gaussian filter to a scalar or a vector field, or letting
the field evolve through a diffusion process can be interpreted as being similar operations.
Indeed, the kernel of the Gaussian filter is:
\begin{equation}\label{cartesianKernel}
G({\bf x} - {\bf x^\prime}) = \left(\frac{\gamma}{\pi \overline{\Delta}^2}  \right)^{\frac{d}{2}}
\exp \left( {-\frac{\gamma |{\bf x} - {\bf x^\prime}|^2}{\overline{\Delta}^2}}\right) \ ,
\end{equation}
where $d$ corresponds to the spatial dimension, $\overline{\Delta}$ is the filter width,
and $\gamma$ is a constant usually set to $6$.
The solution of the diffusion equation:
\begin{equation}\label{diffusion}
\partial_t\xi({\bf x},t) = D \Delta \xi({\bf x},t) \ ,
\end{equation}
where $D$ is a diffusion coefficient and $t$ the time, is the convolution between the scalar field 
$\xi({\bf x^{\prime}},t=0)$, and the Green function:
\begin{equation}\label{greenDiffusion}
G({\bf x} - {\bf x^\prime},t) = \frac{1}{(4 \pi Dt)^{\frac{d}{2}}} 
\exp \left( -\frac{|{\bf x} - {\bf x^\prime}|^2}{4Dt}\right) \ .
\end{equation}
So if one sets $Dt = \frac{\overline{\Delta}^2}{4\gamma}$, diffusing a scalar
field through equation (\ref{diffusion}) is equivalent to filtering that field with the convolution
kernel expressed in (\ref{cartesianKernel}). Based on this observation, \cite{Bulow2004} 
derived a Gaussian-like filter on the surface of a sphere of radius $R$, 
by determining the convolution kernel of the spherical diffusion equation:
\begin{equation}\label{diffusion}
\partial_t\xi({\bf x},t) = D \Delta_{_H} \xi({\bf x},t) \ .
\end{equation}
which reads:
\begin{equation}\label{SHkernel}
G = \sum_{l \epsilon \mathbb{N}} \sqrt{\frac{2l+1}{4\pi}}
Y_{l0} \exp \left( -\frac{l(l+1)Dt}{R^2} \right) \ ,
\end{equation}
where $Y_{l0}$ is the spherical harmonic of degree $l$ and order $m=0$.
So in spectral space, the filtering operation of a scalar field $\xi$ expanded in spherical harmonics:
\begin{equation}
\xi =  \sum_{l \epsilon \mathbb{N}} \sum_{m=-l}^{m=l} \xi_{l,m} Y_{lm}
\end{equation}
simply reduces to the operation:
\begin{equation}\label{spectralFilter}
\overline{\xi}_{l,m} = \xi_{l,m} \exp \left( -\frac{l(l+1)\overline{\Delta}^2}{24 R^2} \right) \ .
\end{equation}

\section{Discretization of the Core Mantle Boundary}\label{appB}
\subsection{Construction of the grid}

The grid describing the Core Mantle Boundary is obtained by recursively subdividing
an initial icosahedron as explained in \cite{Baumgardner1985,Stuhne1999}. For each
grid refinement procedure, a node is added in the middle of the geodesic arc linking 
every two neighboring points. The refinement degree $rd$ of the grid corresponds to 
the number of time this procedure has been applied. Therefore $rd=0$ corresponds to 
the icosahedron itself which possess $N_p=12$ nodes, and $N_c=20$ spherical triangle cells.
As the refinement degree increases, the number of grid points and cells increases as:
\begin{eqnarray}
N_p &=& 2 + 10\times 4^{rd} \label{npoint}\\
N_c &=& 20\times 4^{rd}\label{ncell} \ .
\end{eqnarray}

\noindent To approximate differential operators, a Voronoi-based finite volume
method is chosen. This approach has been widely used in advection-diffusion 
problems such as in \cite{Heikes1994,Lazarov1996,Satoh2008} and has proven to be efficient to 
tackle these kind of problems.
Since in finite volume methods, differential operators are converted to surface integrals, 
control volumes (or Voronoi cells) surrounding each grid points have to be defined. 

\begin{figure}[h!]
\begin{center}
\includegraphics[width=\linewidth]{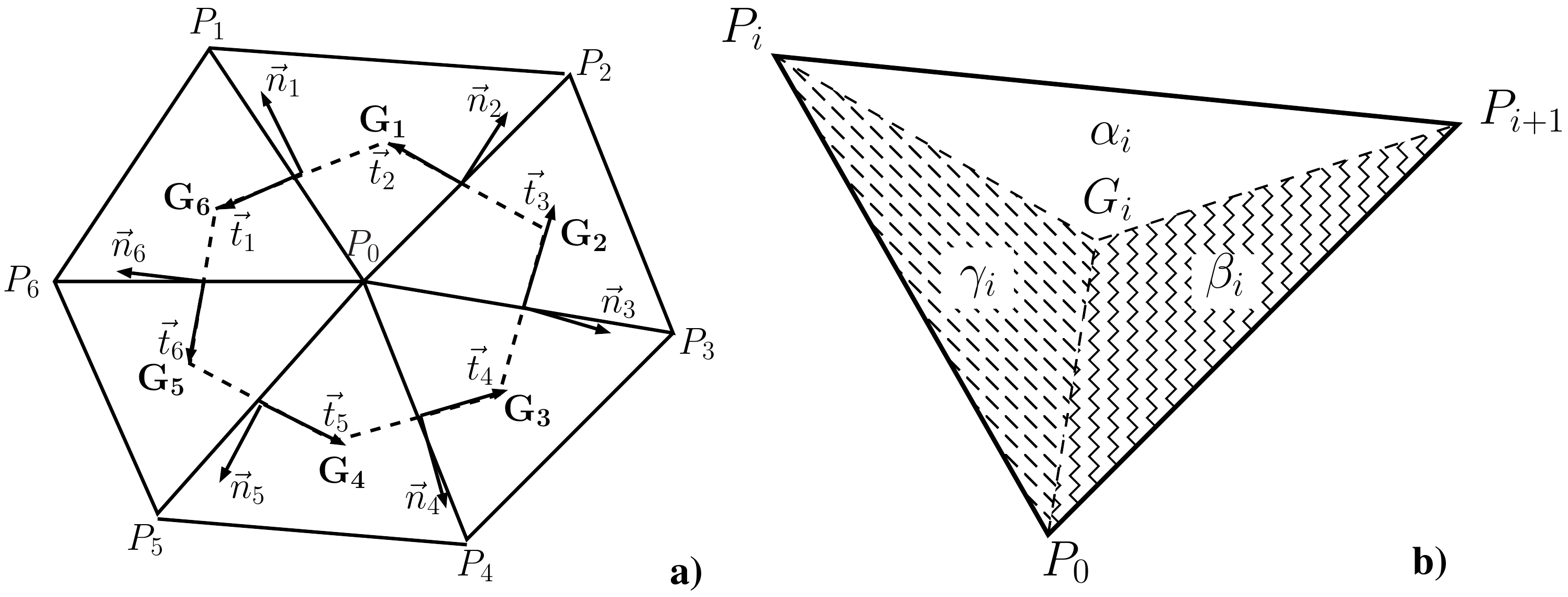}
\caption{a) Voronoi cell delimited by the gravity centers $G_i$ of the different
spherical triangles $P_0,P_i,P_{i+1}$. $\vec{n}_i$ and $\vec{t}_i$ denote respectively the unit
vectors normal and tangential to the voronoi cell contour.  
b) Spherical triangle formed by the three nodes $P_0$, $P_i$ and $P_{i+1}$.
$G_i$ corresponds to the gravity center of the triangle, and $\alpha_i$, $\beta_i$, and $\gamma_i$
are the area of the three sub-triangles.}
\label{operators}
\end{center}
\end{figure}

\noindent As shown in figure \ref{operators} a), each grid point is surrounded by $6$ 
(or $5$ when the point corresponds to a generator of the initial icosahedron) nodes
in its direct neighborhood. From this cluster of node, one can build an ensemble
of spherical triangle cells $P_0 P_i P_{i+1}$ all connected together by the common
vertex $P_0$. Taking the gravity center $G_i$ of each cell allow then to draw a hexagonal 
(or pentagonal) volume control around the central node $P_0$. 

\noindent \cite{Tomita2001,Du2003b} have shown that moving the grid points in a manner
that they coincide with the gravity center of their control volume, allows to improve the
accuracy of the differential operators to the second-order. The procedure employed here is the
Constrained Centroidal Voronoi Tessellations(CCVT) developed by \cite{Du2003a}.
The principle of this procedure as we implemented it, known as the Lloyd's method, 
is the following:

\begin{enumerate}[(1)]
\item starting with an initial distribution of nodes on the sphere,
taken here as the different subdivision of the icosahedron, as explained previously
in this part.

\item building the Voronoi cells associated with each grid point.

\item moving each node to the gravity center of the cell it is belonging to.

\item returning to Step 2 until some convergence criteria is reached. In our case,
we imposed that the averaged geodesic distance between the node and the gravity center 
of the Voronoi cell has to be smaller than $10^{-10}$.
\end{enumerate}

\subsection{Approximation of the differential operators}

The horizontal divergence and curl applied to a vector field ${\bf u}({\bf x})$, 
and the gradient applied to a scalar field $\Phi$ can be expressed in their
integral form as:
\begin{eqnarray}\label{integralOperators}
& {\bf \nabla_{_H}}  \cdot  {\bf u} ({\bf x}) & =  \lim_{\Omega \to 0} \frac{1}{\Omega}
\int_{\delta_\Omega} {\bf u} ({\bf x}) \cdot {\bf t} \; \mathrm{d}\delta_\Omega \\
&{\bf \nabla_{_H}}  \times {\bf u} ({\bf x})&  =  \lim_{\Omega \to 0} \frac{1}{\Omega}
\int_{\delta_\Omega} {\bf u} ({\bf x}) \cdot {\bf n} \; \mathrm{d}\delta_\Omega \\
&{\bf \nabla_{_H}}  \Phi & =  \lim_{\Omega \to 0} \frac{1}{\Omega}
\int_{\delta_\Omega} \Phi  {\bf n} \; \mathrm{d}\delta_\Omega
\end{eqnarray}
where the control surface as an area $\Omega$, delimited by the contour $\delta_\Omega$, 
and the unit tangential and normal vector to the contour are respectively ${\bf t}$ and
${\bf n}$.

\noindent On the discrete sphere, vector and scalar fields are known on the nodes, therefore, to
differentiate them, one need to approximate them on the contour of the Voronoi cells.
First the different quantities are evaluated on the corners of the cells: 
\begin{eqnarray}\label{gravityFields}
{\bf u} (G_i) & = & \frac{\alpha_i {\bf u} (P_0) + \beta_i {\bf u} (P_{i})
+ \gamma_i {\bf u} (P_{i+1})}{\alpha_i  + \beta_i + \gamma_i} \\
\Phi (G_i) & = & \frac{\alpha_i \Phi (P_0) + \beta_i \Phi (P_{i})
+ \gamma_i \Phi (P_{i+1})}{\alpha_i  + \beta_i + \gamma_i} \ ,
\end{eqnarray}
where the areas $\alpha_i$, $\beta_i$, $\gamma_i$ are shown in figure \ref{operators} b).

\noindent Then, following the notation of figure \ref{operators} a) the discrete approximation 
of the different differential operators becomes:
 \begin{eqnarray}\label{approxOperators}
 {\bf \nabla_{_H}}  \cdot  {\bf u} ({\bf x}) & \sim & \frac{1}{A(P_0)} \sum_{i=1}^{N_G}
\widetilde{G_{i} G}_{i+1}
\frac{\left( {\bf u} (G_{i}) + {\bf u} (G_{i+1}  \right)  }{2}\cdot {\bf n}_i\nonumber \\
{\bf \nabla_{_H}}  \times {\bf u} ({\bf x}) & \sim & \frac{1}{A(P_0)} \sum_{i=1}^{N_G}
\widetilde{G_{i} G}_{i+1}
\frac{\left( {\bf u} (G_{i}) + {\bf u} (G_{i+1}  \right)  }{2}\cdot {\bf t}_i \nonumber \\
{\bf \nabla_{_H}}  \Phi &\sim & \sum_{i=1}^{N_G} 
\widetilde{G_{i} G}_{i+1} \frac{1}{A(P_0)} \frac{\left( \Phi (G_{i}) 
+ \Phi (G_{i+1}  \right) }{ 2 } {\bf n}_i \nonumber \\
& & - \frac{\Phi (P_0)}{A(P_0)} \sum_{i=1}^{N_G}
\widetilde{G_{i} G}_{i+1} {\bf n}_i \nonumber
\end{eqnarray}
where $A(P_0)$ corresponds to the area of the control volume, $\widetilde{G_{i} G}_{i+1}$
is the geodesic length between the points $G_{i}$ and $G_{i+1}$, and $N_G$ is the number
of vertices of the control volume. Note that when the subscript $i+1 = N_G +1$ then
$i+1 = 1$.


%

\begin{thebibliography}{43}
\expandafter\ifx\csname natexlab\endcsname\relax\def\natexlab#1{#1}\fi
\expandafter\ifx\csname bibnamefont\endcsname\relax
  \def\bibnamefont#1{#1}\fi
\expandafter\ifx\csname bibfnamefont\endcsname\relax
  \def\bibfnamefont#1{#1}\fi
\expandafter\ifx\csname citenamefont\endcsname\relax
  \def\citenamefont#1{#1}\fi
\expandafter\ifx\csname url\endcsname\relax
  \def\url#1{\texttt{#1}}\fi
\expandafter\ifx\csname urlprefix\endcsname\relax\def\urlprefix{URL }\fi
\providecommand{\bibinfo}[2]{#2}
\providecommand{\eprint}[2][]{\url{#2}}

\bibitem[{\citenamefont{{Vel{\'{\i}}msk{\'y}}}(2010)}]{Velimsky2010}
\bibinfo{author}{\bibfnamefont{J.}~\bibnamefont{{Vel{\'{\i}}msk{\'y}}}},
  \bibinfo{journal}{Physics of the Earth and Planetary Interiors}
  \textbf{\bibinfo{volume}{180}}, \bibinfo{pages}{111} (\bibinfo{year}{2010}).

\bibitem[{\citenamefont{{Roberts} and {Scott}}(1965)}]{Roberts1965}
\bibinfo{author}{\bibfnamefont{P.~H.} \bibnamefont{{Roberts}}}
  \bibnamefont{and} \bibinfo{author}{\bibfnamefont{S.}~\bibnamefont{{Scott}}},
  \bibinfo{journal}{Journal of Geomagnetism and Geoelectricity}
  \textbf{\bibinfo{volume}{17}}, \bibinfo{pages}{137} (\bibinfo{year}{1965}).

\bibitem[{\citenamefont{{Lesur} et~al.}(2010)\citenamefont{{Lesur},
  {Wardinski}, {Hamoudi}, and {Rother}}}]{Lesur2010}
\bibinfo{author}{\bibfnamefont{V.}~\bibnamefont{{Lesur}}},
  \bibinfo{author}{\bibfnamefont{I.}~\bibnamefont{{Wardinski}}},
  \bibinfo{author}{\bibfnamefont{M.}~\bibnamefont{{Hamoudi}}},
  \bibnamefont{and} \bibinfo{author}{\bibfnamefont{M.}~\bibnamefont{{Rother}}},
  \bibinfo{journal}{Earth, Planets, and Space} \textbf{\bibinfo{volume}{62}},
  \bibinfo{pages}{765} (\bibinfo{year}{2010}).

\bibitem[{\citenamefont{{Eymin} and {Hulot}}(2005)}]{Eymin2005}
\bibinfo{author}{\bibfnamefont{C.}~\bibnamefont{{Eymin}}} \bibnamefont{and}
  \bibinfo{author}{\bibfnamefont{G.}~\bibnamefont{{Hulot}}},
  \bibinfo{journal}{Physics of the Earth and Planetary Interiors}
  \textbf{\bibinfo{volume}{152}}, \bibinfo{pages}{200} (\bibinfo{year}{2005}).

\bibitem[{\citenamefont{{Backus}}(1968)}]{Backus1968}
\bibinfo{author}{\bibfnamefont{G.~E.} \bibnamefont{{Backus}}},
  \bibinfo{journal}{Royal Society of London Philosophical Transactions Series
  A} \textbf{\bibinfo{volume}{263}}, \bibinfo{pages}{239}
  (\bibinfo{year}{1968}).

\bibitem[{\citenamefont{{Holme}}(2007)}]{Holme2007}
\bibinfo{author}{\bibfnamefont{R.}~\bibnamefont{{Holme}}}, in
  \emph{\bibinfo{booktitle}{Treatise on Geophysics}}, edited by
  \bibinfo{editor}{\bibfnamefont{E.}~\bibnamefont{in~Chief:Gerald~Schubert}}
  (\bibinfo{publisher}{Elsevier}, \bibinfo{address}{Amsterdam},
  \bibinfo{year}{2007}), pp. \bibinfo{pages}{107 -- 130}, ISBN
  \bibinfo{isbn}{978-0-444-52748-6}.

\bibitem[{\citenamefont{{Finlay} et~al.}(2010)\citenamefont{{Finlay},
  {Dumberry}, {Chulliat}, and {Pais}}}]{Finlay2010}
\bibinfo{author}{\bibfnamefont{C.~C.} \bibnamefont{{Finlay}}},
  \bibinfo{author}{\bibfnamefont{M.}~\bibnamefont{{Dumberry}}},
  \bibinfo{author}{\bibfnamefont{A.}~\bibnamefont{{Chulliat}}},
  \bibnamefont{and} \bibinfo{author}{\bibfnamefont{M.~A.}
  \bibnamefont{{Pais}}}, \bibinfo{journal}{Space Science Reviews}
  \textbf{\bibinfo{volume}{155}}, \bibinfo{pages}{177} (\bibinfo{year}{2010}).

\bibitem[{\citenamefont{{Chulliat} and {Hulot}}(2000)}]{Chulliat2000}
\bibinfo{author}{\bibfnamefont{A.}~\bibnamefont{{Chulliat}}} \bibnamefont{and}
  \bibinfo{author}{\bibfnamefont{G.}~\bibnamefont{{Hulot}}},
  \bibinfo{journal}{Physics of the Earth and Planetary Interiors}
  \textbf{\bibinfo{volume}{117}}, \bibinfo{pages}{309} (\bibinfo{year}{2000}).

\bibitem[{\citenamefont{{Hulot} et~al.}(1992)\citenamefont{{Hulot},
  {Mou{\"e}l}, and {Wahr}}}]{Hulot1992}
\bibinfo{author}{\bibfnamefont{G.}~\bibnamefont{{Hulot}}},
  \bibinfo{author}{\bibfnamefont{J.~L.~L.} \bibnamefont{{Mou{\"e}l}}},
  \bibnamefont{and} \bibinfo{author}{\bibfnamefont{J.}~\bibnamefont{{Wahr}}},
  \bibinfo{journal}{Geophysical Journal International}
  \textbf{\bibinfo{volume}{108}}, \bibinfo{pages}{224} (\bibinfo{year}{1992}).

\bibitem[{\citenamefont{{Gillet} et~al.}(2009)\citenamefont{{Gillet}, {Pais},
  and {Jault}}}]{Gillet2009}
\bibinfo{author}{\bibfnamefont{N.}~\bibnamefont{{Gillet}}},
  \bibinfo{author}{\bibfnamefont{M.~A.} \bibnamefont{{Pais}}},
  \bibnamefont{and} \bibinfo{author}{\bibfnamefont{D.}~\bibnamefont{{Jault}}},
  \bibinfo{journal}{Geochemistry, Geophysics, Geosystems}
  \textbf{\bibinfo{volume}{10}}, \bibinfo{eid}{Q06004} (\bibinfo{year}{2009}).

\bibitem[{\citenamefont{{Pais} and {Jault}}(2008)}]{Pais2008}
\bibinfo{author}{\bibfnamefont{M.~A.} \bibnamefont{{Pais}}} \bibnamefont{and}
  \bibinfo{author}{\bibfnamefont{D.}~\bibnamefont{{Jault}}},
  \bibinfo{journal}{Geophysical Journal International}
  \textbf{\bibinfo{volume}{173}}, \bibinfo{pages}{421} (\bibinfo{year}{2008}).

\bibitem[{\citenamefont{{Hulot} et~al.}(2002)\citenamefont{{Hulot}, {Eymin},
  {Langlais}, {Mandea}, and {Olsen}}}]{Hulot2002}
\bibinfo{author}{\bibfnamefont{G.}~\bibnamefont{{Hulot}}},
  \bibinfo{author}{\bibfnamefont{C.}~\bibnamefont{{Eymin}}},
  \bibinfo{author}{\bibfnamefont{B.}~\bibnamefont{{Langlais}}},
  \bibinfo{author}{\bibfnamefont{M.}~\bibnamefont{{Mandea}}}, \bibnamefont{and}
  \bibinfo{author}{\bibfnamefont{N.}~\bibnamefont{{Olsen}}},
  \bibinfo{journal}{Nature} \textbf{\bibinfo{volume}{416}},
  \bibinfo{pages}{620} (\bibinfo{year}{2002}).

\bibitem[{\citenamefont{{Buffett} and {Christensen}}(2007)}]{Buffett2007}
\bibinfo{author}{\bibfnamefont{B.~A.} \bibnamefont{{Buffett}}}
  \bibnamefont{and} \bibinfo{author}{\bibfnamefont{U.~R.}
  \bibnamefont{{Christensen}}}, \bibinfo{journal}{Geophysical Journal
  International} \textbf{\bibinfo{volume}{171}}, \bibinfo{pages}{145}
  (\bibinfo{year}{2007}).

\bibitem[{\citenamefont{{Roberts} et~al.}(2003)\citenamefont{{Roberts},
  {Jones}, and {Calderwood}}}]{Roberts2003}
\bibinfo{author}{\bibfnamefont{P.~H.} \bibnamefont{{Roberts}}},
  \bibinfo{author}{\bibfnamefont{C.}~\bibnamefont{{Jones}}}, \bibnamefont{and}
  \bibinfo{author}{\bibfnamefont{A.}~\bibnamefont{{Calderwood}}},
  \bibinfo{journal}{Earth's Core and Lower Mantle} pp.
  \bibinfo{pages}{100--129} (\bibinfo{year}{2003}).

\bibitem[{\citenamefont{{Voorhies}}(2004)}]{Voohries2004}
\bibinfo{author}{\bibfnamefont{C.~V.} \bibnamefont{{Voorhies}}},
  \bibinfo{journal}{Journal of Geophysical Research}
  \textbf{\bibinfo{volume}{107}}, \bibinfo{pages}{5034} (\bibinfo{year}{2004}).

\bibitem[{\citenamefont{Sagaut}(2006)}]{Sagaut2006}
\bibinfo{author}{\bibfnamefont{P.}~\bibnamefont{Sagaut}},
  \emph{\bibinfo{title}{Large Eddy Simulation for Incompressible Flows: An
  Introduction}}, Scientific Computation (\bibinfo{publisher}{Springer},
  \bibinfo{year}{2006}), ISBN \bibinfo{isbn}{9783540263449}.

\bibitem[{\citenamefont{{Lesieur}}(2008)}]{Lesieur2008}
\bibinfo{author}{\bibfnamefont{M.}~\bibnamefont{{Lesieur}}},
  \emph{\bibinfo{title}{{Turbulence in Fluids}}}, Fluid Mechanics and Its
  Applications (\bibinfo{publisher}{Springer, Berlin}, \bibinfo{year}{2008}),
  ISBN \bibinfo{isbn}{1402064357, 9781402064357}.

\bibitem[{\citenamefont{{Baerenzung}
  et~al.}(2008{\natexlab{a}})\citenamefont{{Baerenzung}, {Politano}, {Ponty},
  and {Pouquet}}}]{Baerenzung2008a}
\bibinfo{author}{\bibfnamefont{J.}~\bibnamefont{{Baerenzung}}},
  \bibinfo{author}{\bibfnamefont{H.}~\bibnamefont{{Politano}}},
  \bibinfo{author}{\bibfnamefont{Y.}~\bibnamefont{{Ponty}}}, \bibnamefont{and}
  \bibinfo{author}{\bibfnamefont{A.}~\bibnamefont{{Pouquet}}},
  \bibinfo{journal}{Physical Review E} \textbf{\bibinfo{volume}{77}},
  \bibinfo{eid}{046303} (\bibinfo{year}{2008}{\natexlab{a}}),
  \eprint{0707.0642}.

\bibitem[{\citenamefont{{Baerenzung}
  et~al.}(2008{\natexlab{b}})\citenamefont{{Baerenzung}, {Politano}, {Ponty},
  and {Pouquet}}}]{Baerenzung2008b}
\bibinfo{author}{\bibfnamefont{J.}~\bibnamefont{{Baerenzung}}},
  \bibinfo{author}{\bibfnamefont{H.}~\bibnamefont{{Politano}}},
  \bibinfo{author}{\bibfnamefont{Y.}~\bibnamefont{{Ponty}}}, \bibnamefont{and}
  \bibinfo{author}{\bibfnamefont{A.}~\bibnamefont{{Pouquet}}},
  \bibinfo{journal}{Physical Review E} \textbf{\bibinfo{volume}{78}},
  \bibinfo{eid}{026310} (\bibinfo{year}{2008}{\natexlab{b}}),
  \eprint{0803.4499}.

\bibitem[{\citenamefont{{Fabre} and {Balarac}}(2011)}]{Fabre2011}
\bibinfo{author}{\bibfnamefont{Y.}~\bibnamefont{{Fabre}}} \bibnamefont{and}
  \bibinfo{author}{\bibfnamefont{G.}~\bibnamefont{{Balarac}}},
  \bibinfo{journal}{Physics of Fluids} \textbf{\bibinfo{volume}{23}},
  \bibinfo{pages}{115103} (\bibinfo{year}{2011}).

\bibitem[{\citenamefont{{Sun} and {Su}}(2007)}]{Sun2007}
\bibinfo{author}{\bibfnamefont{O.~S.} \bibnamefont{{Sun}}} \bibnamefont{and}
  \bibinfo{author}{\bibfnamefont{L.~K.} \bibnamefont{{Su}}}, in
  \emph{\bibinfo{booktitle}{APS Division of Fluid Dynamics Meeting Abstracts}}
  (\bibinfo{year}{2007}), p.~\bibinfo{pages}{K6}.

\bibitem[{\citenamefont{{B\"ulow}}(2004)}]{Bulow2004}
\bibinfo{author}{\bibfnamefont{T.}~\bibnamefont{{B\"ulow}}},
  \bibinfo{journal}{Transactions on Pattern Analysis and Machine Intelligence}
  \textbf{\bibinfo{volume}{26}} (\bibinfo{year}{2004}).

\bibitem[{\citenamefont{{Driscoll} and {Healy}}(1994)}]{Driscoll1994}
\bibinfo{author}{\bibfnamefont{J.~R.} \bibnamefont{{Driscoll}}}
  \bibnamefont{and} \bibinfo{author}{\bibfnamefont{D.~M.}
  \bibnamefont{{Healy}}}, \bibinfo{journal}{Advances in Applied Mathematics}
  \textbf{\bibinfo{volume}{15}}, \bibinfo{pages}{202 } (\bibinfo{year}{1994}).

\bibitem[{\citenamefont{{Leonard}}(1974)}]{Leonard1974}
\bibinfo{author}{\bibfnamefont{A.}~\bibnamefont{{Leonard}}}, in
  \emph{\bibinfo{booktitle}{Turbulent Diffusion in Environmental Pollution}}
  (\bibinfo{year}{1974}), pp. \bibinfo{pages}{237--248}.

\bibitem[{\citenamefont{{Jackson}}(1995)}]{Jackson1995}
\bibinfo{author}{\bibfnamefont{A.}~\bibnamefont{{Jackson}}},
  \bibinfo{journal}{Physics of the Earth and Planetary Interiors}
  \textbf{\bibinfo{volume}{90}}, \bibinfo{pages}{145} (\bibinfo{year}{1995}).

\bibitem[{\citenamefont{{Bayes}}(1763)}]{Bayes1763}
\bibinfo{author}{\bibfnamefont{T.}~\bibnamefont{{Bayes}}},
  \bibinfo{journal}{Phil. Trans. of the Royal Soc. of London} pp.
  \bibinfo{pages}{370--418} (\bibinfo{year}{1763}).

\bibitem[{\citenamefont{{Bloxham}}(1988)}]{Bloxham1988}
\bibinfo{author}{\bibfnamefont{J.}~\bibnamefont{{Bloxham}}},
  \bibinfo{journal}{Geophysical Research Letters}
  \textbf{\bibinfo{volume}{15}}, \bibinfo{pages}{585} (\bibinfo{year}{1988}).

\bibitem[{\citenamefont{{Jackson} et~al.}(1993)\citenamefont{{Jackson},
  {Bloxham}, and {Gubbins}}}]{Jackson1993}
\bibinfo{author}{\bibfnamefont{A.}~\bibnamefont{{Jackson}}},
  \bibinfo{author}{\bibfnamefont{J.}~\bibnamefont{{Bloxham}}},
  \bibnamefont{and}
  \bibinfo{author}{\bibfnamefont{D.}~\bibnamefont{{Gubbins}}},
  \bibinfo{journal}{Washington DC American Geophysical Union Geophysical
  Monograph Series} \textbf{\bibinfo{volume}{72}}, \bibinfo{pages}{97}
  (\bibinfo{year}{1993}).

\bibitem[{\citenamefont{{Finlay} and {Amit}}(2011)}]{Finlay2011}
\bibinfo{author}{\bibfnamefont{C.~C.} \bibnamefont{{Finlay}}} \bibnamefont{and}
  \bibinfo{author}{\bibfnamefont{H.}~\bibnamefont{{Amit}}},
  \bibinfo{journal}{Geophysical Journal International}
  \textbf{\bibinfo{volume}{186}}, \bibinfo{pages}{175–192}
  (\bibinfo{year}{2011}).

\bibitem[{\citenamefont{{Thebault} and {Vervelidou}}(2013)}]{Thebault2013}
\bibinfo{author}{\bibfnamefont{E.}~\bibnamefont{{Thebault}}} \bibnamefont{and}
  \bibinfo{author}{\bibfnamefont{F.}~\bibnamefont{{Vervelidou}}},
  \bibinfo{journal}{submitted to Physics of the Earth and Planetary Interiors}
  (\bibinfo{year}{2013}).

\bibitem[{\citenamefont{{Mosegaard} and
  {Rygaard-Hjalsted}}(1999)}]{Mosegaard1999}
\bibinfo{author}{\bibfnamefont{K.}~\bibnamefont{{Mosegaard}}} \bibnamefont{and}
  \bibinfo{author}{\bibfnamefont{C.}~\bibnamefont{{Rygaard-Hjalsted}}},
  \bibinfo{journal}{Inverse Problems} \textbf{\bibinfo{volume}{15}},
  \bibinfo{pages}{573} (\bibinfo{year}{1999}).

\bibitem[{\citenamefont{{Rygaard-Hjalsted}
  et~al.}(2000)\citenamefont{{Rygaard-Hjalsted}, {Mosegaard}, and
  Olsen}}]{Rygaard2000}
\bibinfo{author}{\bibfnamefont{C.}~\bibnamefont{{Rygaard-Hjalsted}}},
  \bibinfo{author}{\bibfnamefont{K.}~\bibnamefont{{Mosegaard}}},
  \bibnamefont{and} \bibinfo{author}{\bibfnamefont{N.}~\bibnamefont{Olsen}}, in
  \emph{\bibinfo{booktitle}{Methods and Applications of Inversion}}, edited by
  \bibinfo{editor}{\bibfnamefont{P.}~\bibnamefont{Hansen}},
  \bibinfo{editor}{\bibfnamefont{B.}~\bibnamefont{Jacobsen}}, \bibnamefont{and}
  \bibinfo{editor}{\bibfnamefont{K.}~\bibnamefont{Mosegaard}}
  (\bibinfo{publisher}{Springer Berlin Heidelberg}, \bibinfo{year}{2000}),
  vol.~\bibinfo{volume}{92} of \emph{\bibinfo{series}{Lecture Notes in Earth
  Sciences}}, pp. \bibinfo{pages}{255--275}, ISBN
  \bibinfo{isbn}{978-3-540-65916-7},
  \urlprefix\url{http://dx.doi.org/10.1007/BFb0010296}.

\bibitem[{\citenamefont{{Sukoriansky} et~al.}(2002)\citenamefont{{Sukoriansky},
  {Galperin}, and {Dikovskaya}}}]{Sukoriansky2002}
\bibinfo{author}{\bibfnamefont{S.}~\bibnamefont{{Sukoriansky}}},
  \bibinfo{author}{\bibfnamefont{B.}~\bibnamefont{{Galperin}}},
  \bibnamefont{and}
  \bibinfo{author}{\bibfnamefont{N.}~\bibnamefont{{Dikovskaya}}},
  \bibinfo{journal}{Physical Review Letters} \textbf{\bibinfo{volume}{89}},
  \bibinfo{eid}{124501} (\bibinfo{year}{2002}).

\bibitem[{\citenamefont{{Gamerman} and {Lopes}}(2006)}]{Gamerman2006}
\bibinfo{author}{\bibfnamefont{D.}~\bibnamefont{{Gamerman}}} \bibnamefont{and}
  \bibinfo{author}{\bibfnamefont{H.~F.} \bibnamefont{{Lopes}}},
  \emph{\bibinfo{title}{{Markov Chain Monte Carlo: Stochastic Simulation for
  Bayesian Inference}}}, Chapman and Hall/CRC Texts in Statistical Science
  Series (\bibinfo{publisher}{Taylor \& Francis}, \bibinfo{year}{2006}), ISBN
  \bibinfo{isbn}{9781584885870}.

\bibitem[{\citenamefont{{Lesur} et~al.}(2013)\citenamefont{{Lesur},
  {Wardinski}, and {Whaler}}}]{Lesur2013}
\bibinfo{author}{\bibfnamefont{V.}~\bibnamefont{{Lesur}}},
  \bibinfo{author}{\bibfnamefont{I.}~\bibnamefont{{Wardinski}}},
  \bibnamefont{and} \bibinfo{author}{\bibfnamefont{K.}~\bibnamefont{{Whaler}}},
  in \emph{\bibinfo{booktitle}{IAGA Scientific Assembly}}
  (\bibinfo{year}{2013}).

\bibitem[{\citenamefont{{Baumgardner} and
  {Frederickson}}(1985)}]{Baumgardner1985}
\bibinfo{author}{\bibfnamefont{J.~R.} \bibnamefont{{Baumgardner}}}
  \bibnamefont{and} \bibinfo{author}{\bibfnamefont{P.~O.}
  \bibnamefont{{Frederickson}}}, \bibinfo{journal}{SIAM Journal on Numerical
  Analysis} \textbf{\bibinfo{volume}{22}}, \bibinfo{pages}{1107}
  (\bibinfo{year}{1985}).

\bibitem[{\citenamefont{{Stuhne} and {Peltier}}(1999)}]{Stuhne1999}
\bibinfo{author}{\bibfnamefont{G.~R.} \bibnamefont{{Stuhne}}} \bibnamefont{and}
  \bibinfo{author}{\bibfnamefont{W.~R.} \bibnamefont{{Peltier}}},
  \bibinfo{journal}{Journal of Computational Physics}
  \textbf{\bibinfo{volume}{148}}, \bibinfo{pages}{23} (\bibinfo{year}{1999}).

\bibitem[{\citenamefont{{Heikes} and {Randall}}(1995)}]{Heikes1994}
\bibinfo{author}{\bibfnamefont{R.}~\bibnamefont{{Heikes}}} \bibnamefont{and}
  \bibinfo{author}{\bibfnamefont{D.~A.} \bibnamefont{{Randall}}},
  \bibinfo{journal}{Monthly Weather Review} \textbf{\bibinfo{volume}{123}},
  \bibinfo{pages}{1862} (\bibinfo{year}{1995}).

\bibitem[{\citenamefont{{Lazarov} et~al.}(1996)\citenamefont{{Lazarov},
  {Mishev}, and {Vassilevski}}}]{Lazarov1996}
\bibinfo{author}{\bibfnamefont{R.}~\bibnamefont{{Lazarov}}},
  \bibinfo{author}{\bibfnamefont{P.}~\bibnamefont{{Mishev}}}, \bibnamefont{and}
  \bibinfo{author}{\bibfnamefont{P.}~\bibnamefont{{Vassilevski}}},
  \bibinfo{journal}{Journal of Numerical Analysis}
  \textbf{\bibinfo{volume}{33}}, \bibinfo{pages}{31} (\bibinfo{year}{1996}).

\bibitem[{\citenamefont{{Satoh} et~al.}(2008)\citenamefont{{Satoh}, {Matsuno},
  {Tomita}, {Miura}, {Nasuno}, and {Iga}}}]{Satoh2008}
\bibinfo{author}{\bibfnamefont{M.}~\bibnamefont{{Satoh}}},
  \bibinfo{author}{\bibfnamefont{T.}~\bibnamefont{{Matsuno}}},
  \bibinfo{author}{\bibfnamefont{H.}~\bibnamefont{{Tomita}}},
  \bibinfo{author}{\bibfnamefont{H.}~\bibnamefont{{Miura}}},
  \bibinfo{author}{\bibfnamefont{T.}~\bibnamefont{{Nasuno}}}, \bibnamefont{and}
  \bibinfo{author}{\bibfnamefont{S.}~\bibnamefont{{Iga}}},
  \bibinfo{journal}{Journal of Computational Physics}
  \textbf{\bibinfo{volume}{227}}, \bibinfo{pages}{3486} (\bibinfo{year}{2008}).

\bibitem[{\citenamefont{{Tomita} et~al.}(2001)\citenamefont{{Tomita},
  {Tsugawa}, {Satoh}, and {Goto}}}]{Tomita2001}
\bibinfo{author}{\bibfnamefont{H.}~\bibnamefont{{Tomita}}},
  \bibinfo{author}{\bibfnamefont{M.}~\bibnamefont{{Tsugawa}}},
  \bibinfo{author}{\bibfnamefont{M.}~\bibnamefont{{Satoh}}}, \bibnamefont{and}
  \bibinfo{author}{\bibfnamefont{K.}~\bibnamefont{{Goto}}},
  \bibinfo{journal}{Journal of Computational Physics}
  \textbf{\bibinfo{volume}{174}}, \bibinfo{pages}{579} (\bibinfo{year}{2001}).

\bibitem[{\citenamefont{{Du} et~al.}(2003b)\citenamefont{{Du}, {Gunzburger},
  and {Ju}}}]{Du2003b}
\bibinfo{author}{\bibfnamefont{Q.}~\bibnamefont{{Du}}},
  \bibinfo{author}{\bibfnamefont{M.}~\bibnamefont{{Gunzburger}}},
  \bibnamefont{and} \bibinfo{author}{\bibfnamefont{L.}~\bibnamefont{{Ju}}},
  \bibinfo{journal}{Computational Methods in Applied Mechanics and Engineering}
  \textbf{\bibinfo{volume}{192}}, \bibinfo{pages}{3933}
  (\bibinfo{year}{2003b}).

\bibitem[{\citenamefont{{Du} et~al.}(2003a)\citenamefont{{Du}, {Gunzburger},
  and {Ju}}}]{Du2003a}
\bibinfo{author}{\bibfnamefont{Q.}~\bibnamefont{{Du}}},
  \bibinfo{author}{\bibfnamefont{M.}~\bibnamefont{{Gunzburger}}},
  \bibnamefont{and} \bibinfo{author}{\bibfnamefont{L.}~\bibnamefont{{Ju}}},
  \bibinfo{journal}{Journal of Scientific Computation}
  \textbf{\bibinfo{volume}{24}}, \bibinfo{pages}{1488} (\bibinfo{year}{2003a}).

\end{thebibliography}

\end{document}